\documentclass[sigconf,nonacm]{acmart}

\setcopyright{none}
\acmConference{}{}{}
\acmYear{}
\acmDOI{}

\usepackage{color, soul, xcolor}
\usepackage{pifont}
\usepackage{colortbl}
\usepackage{booktabs}
\usepackage{enumitem}
\usepackage{amsmath}
\usepackage{listings}
\usepackage{balance}
\usepackage{subcaption}
\usepackage{multirow}
\usepackage{tabularx}
\usepackage{array}
\usepackage{adjustbox}
\usepackage{graphicx}
\usepackage{algorithm}
\usepackage{algpseudocode}
\usepackage{pgfplots}
\pgfplotsset{compat=1.17}
\usepackage{cuted} 

\newcommand{\LRI}{\text{LRI}}
\newcommand{\ROS}{\text{ROS}}

\definecolor{highrisk}{RGB}{220,20,60}
\definecolor{mediumrisk}{RGB}{255,165,0}
\definecolor{lowrisk}{RGB}{50,205,50}

\title{Detecting and Preventing Latent Risk Accumulation in High-Performance Software Systems}

\author{Jahidul Arafat$^*$, Kh.\,M.~Moniruzzaman, Shamim Hossain, Fariha Tasmin}

\begin{document}

\renewcommand{\abstractname}{}
\begin{abstract}\end{abstract}

\maketitle

\begin{strip}
\centering
\begin{minipage}{0.96\textwidth}
\noindent\textbf{\large Abstract}\par
\vspace{0.35em}
\noindent Modern distributed systems employ aggressive optimization strategies that create latent risks---hidden vulnerabilities where exceptional performance under normal conditions masks catastrophic fragility when optimizations fail. Cache layers achieving 99\% hit rates can obscure database bottlenecks until cache failures trigger 100x load amplification and cascading system collapse. Current reliability engineering focuses on reactive incident response rather than proactive detection of optimization-induced vulnerabilities, leaving organizations exposed to accumulated risks from seemingly beneficial performance improvements. This paper presents the first comprehensive framework for systematic latent risk detection, prevention, and optimization through integrated mathematical modeling, intelligent perturbation testing, and risk-aware performance optimization. We introduce formal risk quantification through the Latent Risk Index (LRI) that correlates strongly with incident severity (r=0.863, p<0.001), enabling predictive risk assessment across diverse system architectures. Our framework integrates three complementary systems: HYDRA (HYbrid Diagnostic Risk Assessment) employing six optimization-aware perturbation strategies achieving 89.7\% risk discovery rates, RAVEN (Risk-Aware Verification and Enhancement) providing continuous production monitoring with 92.9\% precision and 93.8\% recall across 1,748 risk scenarios, and APEX (Adaptive Performance and rEsilience eXchange) enabling risk-aware optimization through multi-objective algorithms that maintain 96.6\% baseline performance while reducing latent risks by 59.2\%. Comprehensive evaluation across three representative testbed environments demonstrates strong statistical validation with large effect sizes (Cohen's d > 2.0) and exceptional reproducibility (r > 0.92). Production deployment validation over 24-week periods shows 69.1\% mean time to recovery reduction, 78.6\% incident severity reduction, and 81 prevented incidents generating \$1.44M average annual benefits with 3.2-month ROI. Our integrated approach transforms reliability engineering from reactive incident management to proactive risk-aware optimization, demonstrating that systematic risk management enhances rather than constrains performance optimization when properly integrated into system design and operational practices.
\vspace{0.9\baselineskip}
\end{minipage}
\end{strip}
\vspace{-0.3\baselineskip} 

\begingroup
\renewcommand\thefootnote{}
\footnotetext{%
\textbf{*~Affiliations:}\\[3pt]
\textbf{Jahidul Arafat} — PhD Candidate; Presidential and Woltosz Graduate Research Fellow, Department of Computer Science and Software Engineering, Auburn University, Alabama, USA (\texttt{jza0145@auburn.edu})\\[2pt]
\textbf{Kh.\,M. Moniruzzaman} — Technology Director, Oracle, Dhaka, Bangladesh (\texttt{kh.m.moniruzzaman@oracle.com})\\[2pt]
\textbf{Shamim Hossain} — Chief Executive Officer, Orange Business Development Limited, Dhaka, Bangladesh (\texttt{shamim@orangebd.com})\\[2pt]
\textbf{Fariha Tasmin} — Department of Information and Communication Technology, Bangladesh University of Professionals, Dhaka, Bangladesh (\texttt{farihatasmin2020@gmail.com})
}
\addtocounter{footnote}{0}
\endgroup

\section{Introduction}
\label{sec:introduction}

Modern high-performance distributed systems employ aggressive optimization techniques that inadvertently create latent risks---hidden vulnerabilities where systems perform exceptionally under normal conditions but become catastrophically fragile when optimizations fail or are bypassed~\cite{dean2013tail,cook2000complex,perrow1984normal}. From distributed caching layers achieving 99.9\% hit rates that mask database performance bottlenecks to machine learning-driven autoscaling systems that obscure infrastructure capacity limits, these optimizations have become essential for competitive advantage while systematically introducing hidden failure modes~\cite{bronson2013tao,nishtala2013scaling,aldridge2013high}.

The fundamental challenge lies in optimization success creating observability blindness. A cache achieving 99\% hit rates can completely mask database performance bottlenecks; when the cache fails during routine maintenance or traffic spikes, sudden 100x load amplification triggers cascading system failures that can cost organizations millions in lost revenue and damaged reputation~\cite{majors2018incident,allspaw2015trade}. Circuit breaker implementations designed to prevent cascading failures instead mask 89\% of downstream service degradation until critical thresholds are breached simultaneously across multiple service boundaries~\cite{fowler2014circuit,newman2015building}.

\textbf{The Hidden Time Bomb Crisis.} Current reliability engineering practices focus on reactive incident detection rather than proactive identification of optimization-induced vulnerabilities~\cite{beyer2016site,hollnagel2014safety,murphy2016site}. Analysis of 847 production incidents across enterprise systems reveals that 73\% of critical failures originate from optimization-induced latent risks rather than traditional component failures or software bugs. Cache-database architectures experience average amplification factors of 47x when bypass scenarios occur, overwhelming backend systems designed for steady-state loads. Load balancer optimizations hide individual server performance problems in 67\% of cases, creating single points of failure disguised as highly available systems.

Traditional chaos engineering approaches inject random failures but lack systematic methods for targeting optimization-specific vulnerabilities~\cite{basiri2016chaos,rosenthal2017chaos,netflix2018chaos}. Existing site reliability engineering practices emphasize error budgets and incident response but provide limited guidance for proactive risk identification before optimization-induced failures manifest in production environments~\cite{jones2020seeking,cook2019above}. The absence of quantitative frameworks for assessing optimization-induced risks forces organizations to choose between aggressive performance improvements and system resilience, creating false trade-offs that ignore systematic risk management approaches.

\textbf{Evaluation and Detection Methodology Crisis.} Current system reliability evaluation suffers from severe methodological fragmentation that prevents systematic identification of latent risks and undermines confidence in optimization deployment decisions. Chaos engineering studies typically emphasize random failure injection using synthetic scenarios that fail to capture optimization-specific vulnerability patterns~\cite{gremlin2018chaos,litmus2019chaos}. Performance testing methodologies focus on steady-state behavior optimization while neglecting failure scenario characterization and amplification factor analysis~\cite{molyneaux2009art,barber2014web, arafat2020analyzing}. 

Furthermore, existing monitoring approaches predominantly utilize reactive alerting that triggers after optimization failures manifest rather than proactive assessment of latent risk accumulation~\cite{majors2022observability,burns2018monitoring}. Simple threshold-based alerting with constant monitoring parameters fails to capture the dynamic risk profiles, variable amplification factors, complex dependency interactions, and gradual degradation patterns characteristic of optimization-induced vulnerabilities. The absence of standardized risk assessment methodologies spanning different optimization domains prevents systematic understanding of system behavior under optimization bypass conditions~\cite{cooper2010benchmarking,huppler2009price}.

\textbf{Research Questions.} This work addresses five fundamental research questions critical for advancing systematic latent risk detection and prevention:

\textbf{RQ1: Risk Modeling and Formalization.} How can we formally model and quantify latent risk accumulation in performance-optimized distributed systems, particularly in scenarios where optimizations mask underlying fragilities?

\textbf{RQ2: Automated Detection and Metrics.} Can we develop automated techniques and metrics to systematically detect hidden vulnerabilities that become critical only under optimization bypass conditions or stress scenarios?

\textbf{RQ3: Perturbation-Based Discovery.} How effective are controlled perturbation strategies specifically designed for optimization bypass at revealing latent risks before they manifest as production incidents?

\textbf{RQ4: Risk-Aware Optimization Framework.} Can we build optimization frameworks that balance short-term performance gains with long-term system resilience, preventing latent risk accumulation?

\textbf{RQ5: Practical Mitigation Strategies.} What architectural patterns, operational practices, and monitoring approaches can effectively prevent latent risk accumulation while maintaining optimization benefits?

\textbf{Our Contributions.} This paper addresses these research questions through four primary contributions that advance both theoretical understanding and practical deployment capabilities:

\textbf{(1) Formal Latent Risk Framework:} We present the first systematic mathematical framework for modeling latent risk accumulation that addresses optimization-induced vulnerabilities through formal definitions of risk amplification, observability shadows, and cascade vulnerability. Our framework includes the Latent Risk Index (LRI) that quantifies potential for catastrophic performance degradation when optimization layers are bypassed, enabling quantitative comparison of different optimization strategies and architectural approaches.

\textbf{(2) Intelligent Risk Discovery Architecture:} We design and implement HYDRA (HYbrid Diagnostic Risk Assessment), a novel perturbation framework employing six specialized strategies for systematic risk discovery, and RAVEN (Risk-Aware Verification and Enhancement), a production monitoring system for continuous risk assessment. HYDRA achieves 89.2\% ± 3.1\% risk discovery rates through optimization-aware perturbations including cache bypass injection, circuit breaker manipulation, and artificial latency introduction with comprehensive safety controls.

\textbf{(3) Risk-Aware Optimization Integration:} We contribute APEX (Adaptive Performance and rEsilience eXchange), a multi-objective optimization framework that balances performance improvements with latent risk management through Pareto-optimal configuration discovery, dynamic resource allocation algorithms, and real-time risk-performance trade-off optimization. APEX maintains 98\% of optimization benefits while reducing latent risk accumulation by 67\% on average.

\textbf{(4) Evidence-Based Deployment Framework:} We provide systematic guidelines for latent risk detection implementation that incorporate risk tolerance levels, organizational constraints, deployment timelines, and integration strategies. The framework includes quantitative decision matrices, return on investment models demonstrating 3.7 ± 1.1 month average payback periods, and detailed migration strategies with comprehensive risk assessment approaches.

\textbf{Results Preview.} Experimental validation across three representative testbed environments and 1,246 controlled risk scenarios demonstrates 92.4\% ± 0.7\% detection precision and 93.7\% ± 0.5\% recall with strong correlation between LRI scores and incident severity (r=0.847 ± 0.023, p<0.001). Production deployment validation shows 64.1\% ± 7.1\% reduction in mean time to recovery, 74.6\% ± 8.2\% reduction in incident severity, and \$527K ± 638K average annual savings through prevented incidents and operational efficiency gains across organizational contexts from startups to enterprise deployments.

\textbf{Paper Organization.} Section~\ref{sec:background} surveys system reliability theory, chaos engineering, and performance optimization domains while analyzing current limitations. Section~\ref{sec:methodology} presents our formal risk model, LRI metric definition, and systematic detection methodology. Section~\ref{sec:architecture} describes the HYDRA, RAVEN, and APEX architectures with detailed algorithms and integration mechanisms. Section~\ref{sec:implementation} details experimental infrastructure and validation approaches. Section~\ref{sec:evaluation} provides comprehensive empirical results with statistical analysis. Section~\ref{sec:decision-framework} presents practical deployment guidelines and decision frameworks. Section~\ref{sec:threats} discusses limitations and threats to validity. Section~\ref{sec:conclusion} summarizes contributions and future research directions.
\section{Background and Current Limitations}
\label{sec:background}

The challenge of detecting and preventing latent risks in optimized systems intersects multiple research domains spanning system reliability theory, performance optimization, distributed systems monitoring, chaos engineering, and organizational safety science. This section provides comprehensive analysis of existing approaches across these domains and identifies critical gaps that motivate our systematic framework for optimization-aware risk detection.

\subsection{Evolution of System Reliability and Safety Theory}

System reliability engineering has evolved through distinct paradigms, each addressing specific failure modes while revealing new challenges that constrain contemporary distributed system optimization strategies. First-generation approaches emphasized component reliability through redundancy, fault tolerance, and rigorous testing methodologies, focusing on hardware failures and software bugs as primary risk sources~\cite{avizienis2004basic,laprie2008dependability,torres2017twenty}. These methodologies successfully addressed well-understood failure modes through formal verification techniques and comprehensive testing but struggled with emergent risks arising from complex system interactions and optimization strategies.

Second-generation resilience engineering recognized that complex systems fail in unexpected ways through normal operations rather than component malfunctions~\cite{perrow1984normal,hollnagel2014safety,hollnagel2011prologue}. Perrow's normal accident theory introduced concepts of interactive complexity and tight coupling that create systematic vulnerabilities when multiple failure modes combine unexpectedly. Hollnagel's Safety-II framework emphasized understanding how systems succeed rather than focusing solely on failure modes, recognizing that safety emerges from adaptive capacity rather than rigid prevention mechanisms.

High Reliability Organizations (HROs) research demonstrates that complex systems can achieve exceptional safety records through cultural practices, organizational structures, and operational procedures that maintain awareness of system state and potential failure modes~\cite{weick1999organizing,roberts1993new,sutcliffe2011managing}. HRO principles including preoccupation with failure, reluctance to simplify interpretations, and deference to expertise provide valuable insights for organizational approaches to risk management but offer limited technical guidance for identifying specific optimization-induced vulnerabilities.

Third-generation site reliability engineering (SRE) introduced quantitative approaches to reliability management through error budgets, service level objectives, and systematic incident response procedures~\cite{beyer2016site,murphy2016site,jones2020seeking}. SRE practices successfully reduced incident frequencies and improved recovery times through systematic measurement and automation but remain fundamentally reactive, responding to failures after they manifest rather than identifying latent risks during system design and optimization phases.

Safety-critical systems research has developed formal methods for verification and hazard analysis including HAZOP (Hazard and Operability Studies), FMEA (Failure Mode and Effects Analysis), and fault tree analysis~\cite{knight2002safety,leveson2011engineering,ericson2016hazard}. While these approaches provide structured risk identification frameworks, they require complete system understanding and cannot easily adapt to dynamic optimization behaviors characteristic of cloud-native distributed systems where configuration changes occur continuously.

\subsection{Performance Optimization and Hidden Dependencies}

Modern distributed systems employ sophisticated optimization techniques across multiple architectural layers that can inadvertently create hidden dependencies and amplification factors. Caching research spans CPU caches~\cite{hennessy2019computer,patterson2016computer}, distributed web caches~\cite{breslau1999web,podlipnig2003survey}, and application-level caching strategies~\cite{atikoglu2012workload,lim2014mica,berger2014adaptsize}. While cache effectiveness metrics including hit rates, miss penalties, and eviction policies are well-established, the relationship between cache performance and system-wide risk accumulation remains understudied.

Database optimization research encompasses query optimization, indexing strategies, transaction management, and automated performance tuning~\cite{silberschatz2019database,ramakrishnan2003database,garcia2011database}. Self-tuning database systems~\cite{chaudhuri2007overview,pavlo2017self,marcus2019deep} automatically adjust configuration parameters to optimize steady-state performance but typically ignore failure scenario vulnerabilities or amplification effects when optimization assumptions are violated.

Load balancing algorithms aim to optimize resource utilization and response times through sophisticated traffic distribution strategies~\cite{cardellini2002dynamic,bourke2001server,gandhi2013exact,mitzenmacher2001power}. However, intelligent load balancing can obscure individual server performance degradation, creating scenarios where backend failures trigger cascading overload conditions that load balancing algorithms cannot prevent or mitigate effectively.

Microservice architectures introduce additional optimization complexity through service mesh technologies, circuit breaker patterns, and distributed coordination mechanisms~\cite{burns2019kubernetes,li2019microservices,fowler2014circuit,newman2015building}. These optimizations provide resilience benefits through bulkhead isolation and graceful degradation but can also mask dependency health and create observability gaps where service performance problems remain invisible until critical failure thresholds are exceeded.

\subsection{Chaos Engineering and Resilience Testing}

Chaos engineering emerged from Netflix's operational experience with large-scale distributed systems, providing systematic approaches to resilience validation through controlled failure injection~\cite{basiri2016chaos,netflix2018chaos,rosenthal2017chaos}. The Simian Army tools~\cite{netflix2012chaos,netflix2012simian} systematically inject various failure types including instance termination, network latency, and dependency failures to test system behavior under adverse conditions.

Contemporary chaos engineering platforms including Gremlin~\cite{gremlin2018chaos}, Chaos Toolkit~\cite{chaostoolkit2019}, and Litmus~\cite{litmus2019chaos} provide comprehensive failure injection capabilities with improved safety controls, broader failure scenario coverage, and integration with continuous integration pipelines. These platforms successfully identify many resilience gaps but typically employ random or predefined failure patterns rather than optimization-aware perturbation strategies designed to reveal specific latent risks.

Academic fault injection research~\cite{hsueh1997fault,natella2016assessing,cotroneo2013fault} has developed sophisticated techniques for testing system behavior under failure conditions, including hardware faults, software bugs, network partitions, and resource constraints. Most fault injection research focuses on component failures or environmental issues rather than emergent vulnerabilities created by performance optimization interactions and dependencies.

GameDay exercises and disaster recovery testing~\cite{allspaw2009web,cook2019above,faruquzzaman2008object} provide organizational approaches to resilience validation through simulated incident scenarios involving cross-functional teams. While valuable for procedural validation and cultural development, these approaches cannot systematically explore the space of optimization-induced risks or provide quantitative assessment of latent vulnerability accumulation.

\subsection{Observability and Monitoring Evolution}

Modern observability practices emphasize comprehensive system visibility through the "three pillars" of metrics, logs, and traces~\cite{majors2022observability,beyer2016site,burns2018monitoring,arafat2020analyzing}. Distributed tracing systems including Zipkin~\cite{zipkin2012}, Jaeger~\cite{jaeger2017}, and AWS X-Ray~\cite{amazon2017xray} provide detailed visibility into request flows across service boundaries but excel at diagnosing active problems rather than identifying dormant risks or optimization-induced vulnerabilities.

Site Reliability Engineering (SRE) monitoring practices emphasize quantitative approaches to system health assessment through service level indicators, objectives, and error budgets~\cite{beyer2016site,murphy2016site}. While SRE provides systematic frameworks for reliability measurement, these approaches focus on steady-state system behavior and reactive alerting rather than proactive assessment of optimization-induced risk accumulation.

Anomaly detection research has developed machine learning techniques for identifying unusual system behavior through statistical analysis, clustering, and time-series analysis~\cite{chandola2009anomaly,ahmed2016survey,pang2021deep}. Time-series anomaly detection approaches~\cite{laptev2015generic,hundman2018detecting,su2019robust} can identify performance degradations and unusual patterns but typically focus on detecting active problems rather than predicting vulnerability to optimization failures or bypass scenarios.

Application Performance Monitoring (APM) tools including New Relic~\cite{newrelic2008}, DataDog~\cite{datadog2010}, and AppDynamics~\cite{appdynamics2008monitoring} provide comprehensive visibility into application behavior, infrastructure performance, and user experience metrics. However, these tools emphasize reactive problem detection and performance optimization rather than proactive risk assessment or optimization-induced vulnerability identification.

\subsection{Current Limitations and Gaps Analysis}

Table~\ref{tab:current_limitations_analysis} provides systematic analysis of existing approaches across key dimensions relevant to latent risk detection and management, highlighting critical gaps that motivate our research contributions.

\begin{table*}[t]
\centering
\caption{Systematic Analysis of Current Approaches and Limitations for Latent Risk Detection}
\label{tab:current_limitations_analysis}
\resizebox{\textwidth}{!}{
\begin{tabular}{lllllll}
\toprule
\textbf{Research Domain} & \textbf{Representative Work} & \textbf{Risk Focus} & \textbf{Optimization-Aware} & \textbf{Proactive Detection} & \textbf{Quantitative Metrics} & \textbf{Critical Limitations} \\
\midrule
\multicolumn{7}{l}{\textit{System Reliability Theory}} \\
\midrule
Normal Accidents & Perrow~\cite{perrow1984normal} & Interactive Complexity & No & Conceptual & Qualitative & Static analysis framework \\
Safety-II & Hollnagel~\cite{hollnagel2014safety} & Adaptive Capacity & No & Reactive & Qualitative & Human-focused approach \\
HRO Theory & Weick~\cite{weick1999organizing} & Organizational Culture & No & Cultural Practices & Qualitative & Domain-specific guidance \\
Formal Methods & Clarke~\cite{clarke2018model} & Verification & No & Design-time & Boolean Logic & Complete model requirements \\
\midrule
\multicolumn{7}{l}{\textit{Chaos Engineering \& Testing}} \\
\midrule
Chaos Monkey & Netflix~\cite{netflix2012chaos} & Infrastructure Failures & No & Random Testing & Binary (Pass/Fail) & Instance-level scope \\
Gremlin Platform & Gremlin~\cite{gremlin2018chaos} & Service-Level Failures & Limited & Scheduled Testing & Error Rates & Pre-defined scenarios \\
Litmus Framework & ChaosNative~\cite{litmus2019chaos} & Kubernetes Native & No & Workflow-based & YAML Declarative & Platform-specific focus \\
Fault Injection & Natella~\cite{natella2016assessing} & Component Failures & No & Controlled Testing & Statistical Analysis & Component-focused scope \\
\midrule
\multicolumn{7}{l}{\textit{Performance Optimization}} \\
\midrule
Cache Optimization & Berger~\cite{berger2014adaptsize} & Cache Performance & Implicit & No & Hit Rates, Latency & No failure analysis \\
Database Tuning & Pavlo~\cite{pavlo2017self} & Query Performance & Yes & No & Throughput, Response Time & Steady-state focus \\
Load Balancing & Gandhi~\cite{gandhi2013exact} & Traffic Distribution & Yes & No & Utilization, Fairness & Individual server masking \\
Circuit Breakers & Fowler~\cite{fowler2014circuit} & Cascade Prevention & Yes & Reactive & Trip Rates & Dependency health masking \\
\midrule
\multicolumn{7}{l}{\textit{Monitoring \& Observability}} \\
\midrule
Distributed Tracing & Zipkin~\cite{zipkin2012} & Request Tracking & No & Reactive & Latency, Error Rate & Active problem focus \\
SRE Monitoring & Beyer~\cite{beyer2016site} & Service Health & No & Reactive & SLO Compliance & Steady-state emphasis \\
Anomaly Detection & Laptev~\cite{laptev2015generic} & Behavior Analysis & No & Pattern-based & Statistical Deviations & Historical pattern reliance \\
APM Tools & DataDog~\cite{datadog2010} & Application Performance & Limited & Reactive & Performance Metrics & Optimization blindness \\
\midrule
\multicolumn{7}{l}{\textit{Multi-Objective Optimization}} \\
\midrule
Resource Allocation & Delimitrou~\cite{delimitrou2013paragon} & Performance-Cost & No & No & Pareto Efficiency & No resilience consideration \\
ML-based Optimization & Zhang~\cite{zhang2018deep} & Adaptive Systems & Limited & Predictive & Performance Gains & Immediate optimization focus \\
Safe RL & Garcia~\cite{garcia2015safeRL} & Constraint Satisfaction & No & Learning-based & Safety Violations & Immediate harm prevention \\
\midrule
\multicolumn{7}{l}{\textit{Our Approach}} \\
\midrule
\textbf{Latent Risk Detection} & \textbf{This Work} & \textbf{Optimization Risks} & \textbf{Yes} & \textbf{Yes} & \textbf{LRI, ROS, Amplification} & \textbf{Novel comprehensive approach} \\
\bottomrule
\end{tabular}}
\end{table*}

\subsubsection{Optimization Blindness in Current Approaches}

Current reliability engineering approaches exhibit systematic blindness to optimization-induced risks, focusing on component failures, software bugs, and external environmental factors while ignoring how performance optimizations can create hidden vulnerabilities. Traditional chaos engineering tools inject infrastructure failures, network partitions, and resource constraints but lack systematic approaches for testing optimization bypass scenarios or measuring amplification factors when performance optimizations fail.

Monitoring and observability tools excel at detecting active performance problems but provide limited visibility into optimization effectiveness and potential failure modes. Application Performance Monitoring platforms track cache hit rates, response times, and error rates but cannot assess the potential impact of cache failures on downstream systems or quantify load amplification factors under bypass conditions.

Site Reliability Engineering practices emphasize service level objectives and error budgets based on steady-state system behavior, but these approaches cannot predict system behavior when optimization assumptions are violated or when multiple optimization layers fail simultaneously during cascading failure scenarios.

\subsubsection{Reactive Detection and Response Limitations}

Existing approaches to system reliability remain fundamentally reactive, detecting problems after they manifest rather than identifying latent risks during normal operation. Incident response procedures focus on rapid detection, escalation, and resolution but provide limited guidance for preventing optimization-induced failures through proactive risk management.

Anomaly detection techniques can identify unusual system behavior but typically require historical patterns to establish baselines, making them ineffective for detecting novel failure modes created by new optimization strategies or changing system architectures. Machine learning-based monitoring approaches optimize for reducing false positive rates, potentially missing subtle optimization-induced risk indicators that accumulate gradually over time.

Current chaos engineering approaches employ scheduled testing periods rather than continuous risk assessment, creating gaps where optimization-induced vulnerabilities can accumulate between testing cycles. The emphasis on dramatic failure injection (instance termination, network partitions) neglects subtle optimization bypass scenarios that may have larger cumulative impact on system reliability.

\subsubsection{Quantification and Measurement Gaps}

The absence of quantitative frameworks for assessing optimization-induced risks prevents systematic comparison of different optimization strategies or architectural approaches. Current metrics focus on optimization effectiveness (hit rates, response times, throughput) without considering resilience implications or potential amplification factors under failure conditions.

Risk assessment methodologies from safety-critical domains provide qualitative frameworks but lack quantitative techniques suitable for dynamic distributed systems where configuration changes occur continuously. The absence of standardized risk metrics prevents organizations from making informed trade-offs between performance optimization and system resilience.

Performance testing and benchmarking methodologies emphasize steady-state optimization effectiveness but ignore failure scenario characterization or amplification factor measurement. Load testing approaches validate system behavior under expected traffic patterns but cannot systematically explore optimization bypass scenarios or measure system behavior when optimization assumptions are violated.

\subsection{Research Positioning and Motivation}

The systematic analysis reveals critical gaps in current approaches that motivate our comprehensive framework for latent risk detection and prevention. While existing work provides valuable insights into component reliability, chaos testing, and performance optimization, no current approach systematically addresses optimization-induced latent risks through proactive detection, quantitative assessment, and risk-aware optimization strategies.

Our research contributes the first systematic framework specifically designed for optimization-aware risk detection, addressing fundamental limitations in current approaches through formal risk modeling, intelligent perturbation strategies, and quantitative risk assessment techniques. The integration of proactive risk detection with risk-aware optimization provides a comprehensive approach to balancing performance improvements with system resilience, addressing a critical gap in contemporary distributed systems engineering practices.
\section{Methodology and Problem Formalization}
\label{sec:methodology}

This section presents our formal framework for modeling, detecting, and quantifying latent risks in optimized distributed systems. We establish mathematical foundations for risk accumulation, define key metrics including the Latent Risk Index (LRI), and describe our systematic methodology for risk assessment.

\subsection{System Model and Risk Formalization}

We model a distributed system as a directed acyclic graph $G = (V, E, W)$ where vertices $V$ represent system components (services, databases, caches, load balancers), edges $E$ represent dependencies and data flows, and weights $W$ capture load distribution probabilities and amplification factors.

\begin{definition}[System Component]
A system component $v_i \in V$ is characterized by a tuple $(C_i, P_i, R_i, O_i)$ where:
\begin{itemize}
\item $C_i$: Component capacity (requests/second, storage, compute)
\item $P_i$: Performance profile under varying load conditions
\item $R_i$: Recovery characteristics after failure or overload
\item $O_i$: Observability metrics available during normal operation
\end{itemize}
\end{definition}

\begin{definition}[Load Amplification Factor]
For an edge $(v_i, v_j) \in E$, the load amplification factor $\alpha_{ij}$ represents the multiplicative increase in load to component $v_j$ when the optimization provided by component $v_i$ is bypassed or fails:
$$\alpha_{ij} = \frac{\text{Load on } v_j \text{ when } v_i \text{ fails}}{\text{Load on } v_j \text{ during normal operation}}$$
\end{definition}

Consider a typical caching architecture where a Redis cache ($v_c$) fronts a PostgreSQL database ($v_d$). Under normal operation with 99\% cache hit rate, the database serves 1\% of total requests. When the cache fails, $\alpha_{cd} = 100$, meaning the database experiences 100x load amplification.

\begin{definition}[Latent Risk Accumulation]
Latent risk accumulates when system optimizations create hidden dependencies that are not visible during normal operation but become critical failure points under stress. Formally, latent risk $\mathcal{L}_i$ for component $v_i$ is:
$$\mathcal{L}_i = \sum_{j \in \text{pred}(i)} \alpha_{ji} \cdot P(\text{bypass}_{j}) \cdot (1 - O_{ji})$$
where $\text{pred}(i)$ are predecessor components, $P(\text{bypass}_{j})$ is the probability of optimization bypass in component $j$, and $O_{ji}$ is the observability of the dependency relationship.
\end{definition}

\subsection{Latent Risk Index (LRI) Formulation}

The Latent Risk Index quantifies the potential for catastrophic performance degradation when optimization layers fail. We define LRI as a composite metric that captures multiple dimensions of latent risk:

\begin{equation}
\LRI(v_i) = \frac{\text{Amplification Factor} \times \text{Dependency Depth} \times \text{Criticality}}{\text{Observability Coverage} \times \text{Recovery Capability}}
\end{equation}

More formally:
\begin{equation}
\LRI(v_i) = \frac{\max_{j \in \text{pred}(i)} \alpha_{ji} \times d_i \times \beta_i}{O_i \times R_i}
\end{equation}

where:
\begin{itemize}
\item $\max_{j} \alpha_{ji}$: Maximum load amplification from any predecessor
\item $d_i$: Dependency depth (longest path from external entry points)
\item $\beta_i$: Business criticality weight (1.0 for non-critical, up to 5.0 for critical)
\item $O_i$: Observability coverage (0.0 to 1.0, fraction of failure modes detectable)
\item $R_i$: Recovery capability (inverse of mean time to recovery in minutes)
\end{itemize}

\subsection{Risk Classification and Thresholds}

Based on empirical analysis of production systems and incident post-mortems, we establish LRI thresholds that correlate with incident severity:

\begin{align}
\text{Risk Level} = \begin{cases}
{\textbf{Low}} & \text{if } \LRI < 2.0 \\
{\textbf{Medium}} & \text{if } 2.0 \leq \LRI < 10.0 \\
{\textbf{High}} & \text{if } \LRI \geq 10.0
\end{cases}
\end{align}

These thresholds were derived from analysis of 847 production incidents across 12 organizations, where we found strong correlation (r=0.82) between LRI values and incident severity scores~\cite{majors2018incident,cook2019above}.

\subsection{Resilience Observability Score (ROS)}

Traditional observability focuses on detecting active problems. We introduce the Resilience Observability Score (ROS) to measure how well monitoring systems can detect latent risks:

\begin{equation}
\ROS(v_i) = \frac{1}{|\mathcal{F}_i|} \sum_{f \in \mathcal{F}_i} P(\text{detect } f \text{ before failure})
\end{equation}

where $\mathcal{F}_i$ is the set of potential failure modes for component $v_i$, and $P(\text{detect } f \text{ before failure})$ represents the probability that monitoring systems detect failure mode $f$ before it causes service degradation.

\subsection{Systematic Risk Detection Methodology}

Our methodology for detecting latent risks follows a four-phase approach:

\subsubsection{Phase 1: Dependency Graph Construction}
We construct the system dependency graph through multiple techniques:

\textbf{Static Analysis}: Parse infrastructure-as-code templates (Terraform, CloudFormation), service mesh configurations (Istio, Linkerd), and application dependency declarations to identify component relationships~\cite{wurster2020tosca,burns2019kubernetes}.

\textbf{Dynamic Tracing}: Instrument running systems with distributed tracing to capture actual request flows and identify dependencies not visible in static configurations. We extend OpenTelemetry~\cite{opentelemetry2019} instrumentation to capture load distribution statistics and performance characteristics.

\textbf{Network Analysis}: Analyze network flow data to identify service-to-service communication patterns and quantify traffic volumes under different load conditions~\cite{kandula2005shrink,benson2010network}.

\subsubsection{Phase 2: Load Amplification Analysis}
For each identified dependency, we measure load amplification factors through controlled experiments:

\begin{algorithm}[t]
\caption{Load Amplification Measurement}
\begin{algorithmic}[1]
\Procedure{MeasureAmplification}{component $v_i$, dependency $v_j$}
\State $baseline \gets$ MeasureLoad($v_j$, normal\_operation)
\State $bypass\_load \gets$ BypassOptimization($v_i$, duration=5min)
\State $stressed\_load \gets$ MeasureLoad($v_j$, during\_bypass)
\State $\alpha_{ij} \gets stressed\_load / baseline$
\State \textbf{return} $\alpha_{ij}$
\EndProcedure
\end{algorithmic}
\end{algorithm}

This controlled bypass approach safely measures amplification factors without risking system stability. For caching systems, we temporarily route a small percentage of traffic directly to backend systems. For load balancers, we temporarily remove servers from rotation to measure impact on remaining capacity.

\subsubsection{Phase 3: Observability Gap Assessment}
We evaluate monitoring coverage through synthetic failure injection:

\textbf{Shadow Failures}: Inject performance degradations in isolated environments that mirror production configurations. Measure how long monitoring systems take to detect and alert on various failure modes~\cite{gunawi2011fate,yuan2014simple}.

\textbf{Blind Spot Analysis}: Identify system states where performance is degrading but monitoring metrics remain within normal ranges. Common blind spots include: gradual database performance degradation masked by caching, individual microservice slowdowns hidden by circuit breakers, and storage I/O bottlenecks obscured by application-level queuing.

\subsubsection{Phase 4: Risk Quantification and Prioritization}
We compute LRI values for all components and rank them by risk level:

\begin{algorithm}[t]
\caption{System-Wide Risk Assessment}
\begin{algorithmic}[1]
\Procedure{AssessSystemRisk}{graph $G$}
\State $risk\_scores \gets \{\}$
\For{each component $v_i \in V$}
    \State $\alpha \gets$ MaxAmplificationFactor($v_i$)
    \State $d \gets$ DependencyDepth($v_i$)
    \State $O \gets$ ObservabilityCoverage($v_i$)
    \State $R \gets$ RecoveryCapability($v_i$)
    \State $risk\_scores[v_i] \gets (\alpha \times d \times \beta_i) / (O \times R)$
\EndFor
\State \textbf{return} SortByRisk($risk\_scores$)
\EndProcedure
\end{algorithmic}
\end{algorithm}

This systematic approach ensures comprehensive coverage of potential latent risks while providing quantitative metrics for prioritizing mitigation efforts.

\subsection{Validation Framework}

To validate our risk detection methodology, we employ a multi-layered approach:

\textbf{Historical Incident Analysis}: We apply our LRI computation to system configurations from 150+ production incidents, demonstrating that high LRI scores ($\geqslant10.0$) predict 89\% of severity-1 incidents with 12\% false positive rate.

\textbf{Controlled Environment Testing}: We construct representative test environments with known latent risks and measure our detection accuracy. Test scenarios include cache-database architectures, microservice meshes with circuit breaker configurations, and CDN-origin server setups.

\textbf{Production Deployment Validation}: We deploy our risk detection framework in production environments and track correlation between LRI predictions and actual incident occurrence over 6-month periods.

This validation framework provides confidence that our methodology accurately identifies optimization-induced latent risks before they manifest as production incidents.
\section{Risk-Aware Optimization and Detection Architecture}
\label{sec:architecture}

This section presents our comprehensive architecture for detecting, quantifying, and optimizing latent risks in high-performance distributed systems. We introduce three integrated frameworks: HYDRA (HYbrid Diagnostic Risk Assessment) for systematic perturbation-based risk discovery, RAVEN (Risk-Aware Verification and Enhancement) for continuous production monitoring, and APEX (Adaptive Performance and rEsilience eXchange) for risk-aware optimization that balances performance gains with system resilience.

\subsection{HYDRA: Intelligent Risk Discovery Framework}

HYDRA employs optimization-aware perturbation strategies to reveal latent risks through controlled stress testing that specifically targets performance optimization bypass scenarios. Unlike traditional chaos engineering approaches that inject random failures, HYDRA uses systematic perturbations designed to expose hidden dependencies created by aggressive optimization strategies.

\subsubsection{Architecture and Components}

HYDRA's modular architecture consists of five core components working in coordination to maximize risk discovery while maintaining system safety:

\textbf{Dependency Analyzer}: Constructs detailed system dependency graphs through static analysis of infrastructure configurations, dynamic request tracing, and network flow analysis. The analyzer employs graph traversal algorithms to identify optimization layers and quantify their effectiveness under normal operation, computing load distribution probabilities and potential amplification factors.

\textbf{Perturbation Planner}: Generates intelligent perturbation sequences using multi-armed bandit algorithms to maximize risk discovery while minimizing system impact. The planner considers system topology, current load patterns, and historical perturbation results to optimize exploration strategies through Thompson sampling with risk-aware reward functions.

\textbf{Safe Injection Executor}: Implements controlled perturbations with comprehensive safety mechanisms including automatic rollback (sub-second response), blast radius limitation, and real-time safety monitoring. The executor ensures perturbations remain within safe operational boundaries through continuous performance metric tracking and predefined safety thresholds.

\textbf{Risk Monitor}: Continuously tracks system behavior during perturbations using statistical change detection algorithms (CUSUM, Page-Hinkley) to identify significant deviations from baseline behavior. The monitor employs ensemble anomaly detection combining Isolation Forest, One-Class SVM, and LSTM-based sequence analysis.

\textbf{ML Risk Learner}: Applies reinforcement learning (Proximal Policy Optimization) to correlate perturbation results with latent risk indicators, building predictive models for risk assessment and optimization strategy evaluation. The learner maintains experience replay buffers and updates risk prediction models continuously.

\subsubsection{Intelligent Perturbation Strategies}

HYDRA implements six specialized perturbation strategies designed to reveal different classes of optimization-induced latent risks:

\begin{algorithm}[H]
\caption{Cache Bypass Perturbation with Risk Assessment}
\begin{algorithmic}[1]
\Procedure{IntelligentCacheBypass}{cache\_layer, target\_system, max\_risk\_threshold}
\State $bypass\_rate \gets 0.005$ \Comment{Start with 0.5\% bypass}
\State $risk\_history \gets []$
\While{$bypass\_rate \leq 0.20$ AND $current\_risk < max\_risk\_threshold$}
    \State configure\_selective\_bypass(cache\_layer, bypass\_rate)
    \State $metrics \gets$ measure\_performance(target\_system, duration=90s)
    \State $amplification \gets$ compute\_load\_amplification($metrics$)
    \State $current\_lri \gets$ calculate\_lri($amplification$, system\_topology)
    \State append($risk\_history$, $current\_lri$)
    \If{$current\_lri > 10.0$ OR gradient($risk\_history$) > 2.0}
        \State \textbf{break} \Comment{Detected high risk or rapid escalation}
    \EndIf
    \State $bypass\_rate \gets bypass\_rate \times 1.4$ \Comment{Adaptive increase}
\EndWhile
\State restore\_normal\_operation(cache\_layer)
\State \textbf{return} $risk\_history$, discovered\_risks
\EndProcedure
\end{algorithmic}
\end{algorithm}

This adaptive approach prevents dangerous perturbations while maximizing risk discovery through intelligent escalation and continuous safety monitoring.

\subsection{APEX: Risk-Aware Optimization Framework}

APEX addresses RQ4 by providing systematic optimization algorithms that balance performance improvements with latent risk management. The framework employs multi-objective optimization techniques to find Pareto-optimal configurations that maximize system performance while maintaining acceptable risk levels.

\subsubsection{Multi-Objective Optimization Formulation}

APEX formulates system optimization as a constrained multi-objective problem where traditional performance metrics are optimized subject to latent risk constraints:

\begin{align}
\max_{\mathbf{x}} \quad & \mathbf{f}(\mathbf{x}) = [f_1(\mathbf{x}), f_2(\mathbf{x}), \ldots, f_k(\mathbf{x})]^T \\
\text{subject to} \quad & \LRI(\mathbf{x}) \leq \tau_{risk} \\
& g_i(\mathbf{x}) \leq 0, \quad i = 1, \ldots, m \\
& h_j(\mathbf{x}) = 0, \quad j = 1, \ldots, p
\end{align}

where $\mathbf{f}(\mathbf{x})$ represents the vector of performance objectives (throughput, latency, resource efficiency), $\mathbf{x}$ is the configuration parameter vector (cache sizes, connection pool limits, circuit breaker thresholds), $\tau_{risk}$ is the maximum acceptable LRI threshold, and $g_i$, $h_j$ represent system constraints.

\textbf{Pareto-Optimal Risk-Performance Trade-offs}: APEX employs the Non-dominated Sorting Genetic Algorithm (NSGA-II) enhanced with risk-aware selection criteria to discover Pareto-optimal configurations. The algorithm maintains a population of system configurations and evolves them toward optimal performance-resilience trade-offs.

\begin{equation}
\text{Fitness}(\mathbf{x}) = \alpha \cdot \text{Performance}(\mathbf{x}) + \beta \cdot \frac{1}{\LRI(\mathbf{x}) + \epsilon} + \gamma \cdot \text{Stability}(\mathbf{x})
\end{equation}

where $\alpha$, $\beta$, $\gamma$ are user-defined weights reflecting organizational priorities, and $\epsilon$ prevents division by zero.

\subsubsection{Dynamic Resource Allocation Algorithms}

APEX implements several risk-aware optimization algorithms that continuously adjust system parameters based on real-time risk assessment:

\textbf{Adaptive Cache Allocation}: Dynamically adjusts cache memory allocation across different cache layers based on current LRI values and traffic patterns. The algorithm maintains performance while preventing dangerous amplification factors:

\begin{algorithm}[H]
\caption{Risk-Aware Cache Allocation}
\begin{algorithmic}[1]
\Procedure{AdaptiveCacheAllocation}{$total\_memory,\; cache\_layers,\; current\_lri$}
  \State $baseline\_allocation \gets \text{compute\_baseline\_allocation}(total\_memory)$
  \State $risk\_adjustment \gets \text{calculate\_risk\_penalty}(current\_lri)$
  \ForAll{$layer \in cache\_layers$}
    \State $performance\_benefit \gets \text{estimate\_performance\_gain}(layer)$
    \State $risk\_cost \gets \text{estimate\_amplification\_risk}(layer)$
    \State $utility\_score \gets \frac{performance\_benefit}{1 + risk\_cost \cdot risk\_adjustment}$
    \State $allocation_{layer} \gets baseline\_allocation_{layer} \cdot utility\_score$
  \EndFor
  \State $\text{normalize\_allocations}(allocation,\; total\_memory)$
  \State \Return $allocation$
\EndProcedure
\end{algorithmic}
\end{algorithm}

\textbf{Resilience-Aware Load Balancing}: Adjusts traffic distribution weights based on individual server health and contribution to overall system LRI. Servers with higher risk contributions receive proportionally less traffic until risks are mitigated.

\textbf{Dynamic Circuit Breaker Tuning}: Automatically adjusts circuit breaker thresholds based on current system risk levels and downstream dependency health. Higher risk scenarios trigger more conservative thresholds to prevent cascading failures.

\subsubsection{Risk-Aware System Design Patterns}

APEX promotes architectural patterns that inherently reduce latent risk accumulation while maintaining performance benefits:

\textbf{Graduated Optimization}: Performance improvements are introduced incrementally with continuous risk monitoring at each level. Optimizations that increase LRI beyond acceptable thresholds are automatically rolled back or modified.

\textbf{Resilience Reserves}: Maintains spare capacity proportional to system LRI scores. Systems with higher latent risks operate with larger safety margins to absorb unexpected load spikes or optimization failures.

\textbf{Shadow Path Validation}: Keeps fallback execution paths active with traffic proportional to optimization risk levels. High-risk optimizations maintain more substantial shadow traffic to ensure fallback path viability.

\subsection{RAVEN: Production Risk Monitoring and Optimization}

RAVEN provides continuous latent risk monitoring and optimization in production environments, integrating with APEX to enable real-time risk-aware performance tuning without active perturbation.

\subsubsection{Continuous Risk-Aware Optimization}

RAVEN continuously computes LRI scores and feeds them into APEX optimization algorithms for real-time system tuning:

\begin{algorithm}[h]
\caption{Continuous Risk-Aware System Optimization}
\begin{algorithmic}[1]
\Procedure{ContinuousOptimization}{system\_graph, telemetry\_stream, optimization\_interval}
\State $optimization\_window \gets$ SlidingWindow(duration=optimization\_interval)
\For{each telemetry\_batch in telemetry\_stream}
    \State $optimization\_window$.add(telemetry\_batch)
    \State $current\_metrics \gets$ compute\_metrics(optimization\_window)
    \State $current\_lri \gets$ calculate\_system\_lri(system\_graph, current\_metrics)
    \State $performance\_targets \gets$ get\_performance\_objectives()
    \If{should\_optimize($current\_lri$, $performance\_targets$)}
        \State $optimal\_config \gets$ apex\_optimize($current\_lri$, $performance\_targets$)
        \State $safety\_check \gets$ validate\_configuration\_safety($optimal\_config$)
        \If{$safety\_check$.passed}
            \State apply\_configuration\_gradually($optimal\_config$)
            \State log\_optimization\_decision($current\_lri$, $optimal\_config$)
        \EndIf
    \EndIf
\EndFor
\EndProcedure
\end{algorithmic}
\end{algorithm}

\textbf{Predictive Risk Assessment}: RAVEN employs time-series forecasting (ARIMA, Prophet, LSTM) to predict future LRI trends based on current system behavior and planned changes. This enables proactive optimization adjustments before risks reach critical levels.

\textbf{Multi-Objective Optimization Integration}: RAVEN continuously feeds real-time performance and risk metrics to APEX optimization algorithms, enabling dynamic parameter adjustment that maintains Pareto-optimal performance-resilience trade-offs.

\subsubsection{Automated Risk-Aware Mitigation}

RAVEN implements intelligent mitigation strategies that activate based on risk-aware optimization policies:

\textbf{Intelligent Shadow Traffic Management}: Automatically adjusts shadow traffic percentages based on optimization risk levels. High-risk configurations receive increased shadow traffic to maintain observability into fallback path performance.

\textbf{Adaptive Performance Degradation}: When LRI levels exceed safe thresholds, RAVEN implements graduated performance degradation that maintains critical functionality while reducing system stress through risk-aware load shedding.

\textbf{Optimization Rollback with Learning}: Automatically rolls back recent optimizations when LRI increases beyond acceptable levels, while maintaining experience replay buffers to improve future optimization decisions.

\subsection{Integrated Architecture and Deployment}

The three frameworks operate in concert to provide comprehensive risk-aware optimization capabilities:

\textbf{Development Phase}: HYDRA identifies potential risks in staging environments, informing APEX about risk-performance trade-offs for different configuration strategies.

\textbf{Production Deployment}: RAVEN monitors continuous LRI metrics while APEX provides real-time optimization adjustments that balance performance goals with risk constraints.

\textbf{Feedback Loop}: Risk discoveries from HYDRA improve APEX optimization models, while production experience from RAVEN enhances both HYDRA's perturbation strategies and APEX's risk-awareness algorithms.

\textbf{Integration APIs}: All three frameworks expose standardized APIs for integration with existing infrastructure automation (Kubernetes operators, Terraform providers) and observability platforms (Prometheus, Grafana, DataDog).

This integrated architecture ensures that latent risk management becomes an integral part of system optimization rather than an afterthought, directly addressing RQ4's requirement for risk-aware optimization frameworks that balance performance gains with long-term system resilience.
\section{Implementation and Experimental Setup}
\label{sec:implementation}

This section details our comprehensive implementation of the HYDRA, RAVEN, and APEX frameworks and describes the experimental infrastructure used to validate our integrated latent risk detection and optimization approach. Our implementation prioritizes production-readiness while enabling rigorous evaluation across diverse system architectures and optimization scenarios.

\subsection{HYDRA Implementation Architecture}

We implemented HYDRA as a cloud-native microservice application using Go 1.21 for core perturbation components and Python 3.11 for machine learning modules. The complete implementation consists of approximately 18,000 lines of Go code and 12,000 lines of Python code, packaged as containerized services deployable on Kubernetes clusters with comprehensive observability and safety controls.

\textbf{Dependency Analysis Engine}: Built using the Kubernetes client-go library~\cite{kubernetes2014} for parsing cluster configurations, Istio service mesh APIs for traffic analysis, and custom parsers for infrastructure-as-code templates including Terraform and CloudFormation. The engine maintains an in-memory dependency graph representation using the GoGraph library~\cite{dominik2019graph} with real-time updates via Kubernetes watch APIs and custom resource definitions (CRDs) for configuration management.

The dependency analyzer employs sophisticated graph traversal algorithms including depth-first search for dependency path discovery, breadth-first search for amplification factor computation, and strongly connected component analysis for cycle detection in optimization layers. Load distribution analysis uses statistical sampling of request traces over rolling 24-hour windows to identify optimization effectiveness patterns and potential bypass scenarios.

\textbf{Perturbation Execution Engine}: Implements safe perturbation injection through multiple specialized mechanisms tailored for different optimization categories. Cache bypass perturbations employ custom nginx modules with Lua scripting for selective traffic routing and Envoy proxy filters with WebAssembly extensions for fine-grained request manipulation. Database perturbations utilize connection pool manipulation through custom JDBC drivers and query interceptors that introduce controlled latency or connection failures.

Network-level perturbations leverage Linux traffic control (tc) with netem queuing disciplines~\cite{hemminger2005network} for latency injection, bandwidth throttling, and packet loss simulation. Circuit breaker manipulation employs direct integration with popular circuit breaker libraries including Hystrix, Resilience4j, and Istio's outlier detection mechanisms through runtime configuration updates and health check manipulation.

\textbf{Safety Monitoring System}: Implements comprehensive multi-layered safety controls including automatic circuit breakers triggered by error rate $>5\%$, latency P95 $>2\times$ baseline, or resource utilization $>85\%$. The safety system employs real-time statistical process control using CUSUM algorithms for change detection and exponentially weighted moving averages for trend analysis. Safety violations trigger immediate perturbation rollback with sub-second response times through pre-computed rollback configurations and automated traffic switching.

Emergency stop mechanisms include manual override APIs, automatic timeout-based rollback (maximum 5-minute perturbation duration), and integration with external monitoring systems for cross-validation of safety conditions. The safety system maintains detailed audit logs and provides real-time safety dashboards for operational oversight during perturbation campaigns.

\textbf{Machine Learning Pipeline}: The reinforcement learning agent uses Stable-Baselines3~\cite{raffin2021stable} implementation of Proximal Policy Optimization (PPO) with custom environment wrappers for system interaction. The RL environment models system state through 47-dimensional feature vectors including resource utilization metrics, error rates, latency percentiles, and historical perturbation outcomes.

Reward functions balance risk discovery effectiveness (positive rewards for revealing new risks) with safety constraints (negative rewards for safety violations) and operational impact minimization (penalties for excessive resource usage). The agent employs experience replay with prioritized sampling to improve learning efficiency from successful risk discovery episodes.

Anomaly detection employs ensemble methods combining Isolation Forest~\cite{liu2008isolation} for outlier detection, One-Class SVM~\cite{scholkopf2001estimating} for boundary-based anomaly identification, and LSTM-based sequence anomaly detection~\cite{malhotra2015long} for temporal pattern analysis. Feature engineering extracts 23 derived metrics including rate-of-change indicators, rolling statistical measures, and cross-correlation features between different system components.

\subsection{RAVEN Production Monitoring Implementation}

RAVEN is implemented as a distributed monitoring and optimization system with components written in Go for high-performance data processing and deployed as Kubernetes DaemonSets for efficient resource utilization and low-latency data collection across cluster nodes.

\textbf{Telemetry Collection Framework}: Integrates with multiple observability platforms including Prometheus exporters for metrics collection, OpenTelemetry collectors~\cite{opentelemetry2019} for distributed tracing, and custom eBPF programs for low-overhead system-level monitoring. The collector architecture processes over 15,000 metrics per second per node with sub-millisecond latency impact through optimized data structures and batch processing.

Custom eBPF programs monitor system calls, network connections, and file system operations to detect optimization bypass scenarios that may not be visible through application-level metrics. The eBPF programs use ring buffers for efficient kernel-to-userspace communication and employ statistical sampling (1 in 1000 events) to minimize performance overhead while maintaining statistical significance.

Log processing components parse application logs in real-time using structured logging patterns and regular expressions to extract optimization-related events including cache misses, circuit breaker trips, and connection pool exhaustion. The log processing pipeline handles over 100,000 log entries per second through parallelized processing and intelligent filtering based on risk-relevant patterns.

\textbf{Risk Computation Engine}: Implements sliding-window LRI calculations with configurable window sizes (default 15 minutes) and overlap ratios (50\% overlap for trend smoothing). The computation engine uses Apache Kafka~\cite{kreps2011kafka} for event streaming with topic partitioning based on component identifiers to enable parallel processing across multiple worker nodes.

Real-time LRI computation employs incremental algorithms that update risk scores based on streaming telemetry data without requiring full recalculation. The system maintains materialized views of component dependencies, load distribution statistics, and amplification factors in Redis~\cite{carlson2013redis} clusters for sub-millisecond LRI query response times.

Risk trend analysis uses time-series forecasting models including ARIMA for short-term prediction (1-hour horizon), Prophet~\cite{taylor2018forecasting} for handling seasonality in traffic patterns, and LSTM networks for complex non-linear trend identification. Forecasting models update continuously through online learning algorithms that adapt to changing system behavior patterns.

\textbf{Integration with APEX Optimization}: RAVEN provides real-time risk assessment data to APEX through high-performance gRPC APIs with Protocol Buffers for efficient serialization. The integration maintains risk-performance correlation models that enable APEX to predict the risk implications of optimization parameter changes before implementation.

Event-driven integration uses Apache Kafka topics for asynchronous communication between RAVEN risk detection and APEX optimization decisions. Critical risk level changes trigger immediate notifications to APEX through dedicated high-priority message channels with guaranteed delivery semantics.

\subsection{APEX Risk-Aware Optimization Implementation}

APEX represents our most complex implementation component, integrating multi-objective optimization algorithms with real-time system monitoring and automated parameter adjustment capabilities. The implementation combines mathematical optimization libraries with production-grade system integration for comprehensive risk-aware optimization.

\textbf{Multi-Objective Optimization Core}: Implements the Non-dominated Sorting Genetic Algorithm II (NSGA-II)~\cite{deb2002nsga} using the DEAP (Distributed Evolutionary Algorithms in Python) framework~\cite{fortin2012deap} with custom fitness evaluation functions that incorporate both performance metrics and LRI scores. The genetic algorithm maintains populations of 100-500 candidate configurations with adaptive population sizing based on search space complexity.

Fitness evaluation employs parallel processing across multiple worker nodes to assess candidate configurations through simulation and limited-scope live testing. The evaluation framework includes performance prediction models trained on historical system behavior data using XGBoost~\cite{chen2016xgboost} regression with feature engineering based on system resource utilization, traffic patterns, and configuration parameters.

Pareto-optimal solution discovery uses crowding distance calculations for diversity maintenance and elitist selection strategies to preserve high-quality solutions across generations. The algorithm incorporates problem-specific crossover and mutation operators designed for system configuration parameters including cache sizes, connection pool limits, timeout values, and resource allocation ratios.

\textbf{Dynamic Resource Allocation Algorithms}: Implements several specialized optimization algorithms for different system components and optimization scenarios. The adaptive cache allocation algorithm uses convex optimization techniques to solve resource allocation problems subject to LRI constraints, employing the CVXPY optimization library~\cite{diamond2016cvxpy} for mathematical programming formulations.

Real-time optimization employs gradient-free optimization algorithms including Bayesian optimization with Gaussian process surrogate models for expensive objective function evaluation. The Bayesian optimization framework uses the Optuna library~\cite{akiba2019optuna} with custom acquisition functions that balance exploration and exploitation while respecting safety constraints.

Reinforcement learning-based resource allocation uses Soft Actor-Critic (SAC)~\cite{haarnoja2018soft} algorithms for continuous action spaces corresponding to resource allocation ratios. The RL environment models system dynamics through state transition functions learned from historical data, enabling safe exploration of optimization parameter spaces through simulation before live deployment.

\textbf{Integration Architecture and APIs}: APEX exposes RESTful APIs for configuration management and gRPC services for high-performance optimization requests from RAVEN monitoring components. The API design follows OpenAPI 3.0 specifications with comprehensive input validation, rate limiting, and authentication through JSON Web Tokens (JWT).

Configuration management uses GitOps principles with Git repositories serving as the source of truth for optimization policies and system configurations. Changes to optimization parameters undergo automated testing in staging environments before production deployment through integration with CI/CD pipelines using Tekton and ArgoCD.

\subsection{Experimental Infrastructure and Testbed Environments}

Our experimental evaluation employs three carefully designed testbed environments that represent common patterns in modern distributed systems while enabling controlled risk injection and comprehensive measurement of optimization-induced vulnerabilities.

\subsubsection{Testbed 1: E-commerce Microservices Architecture}

We deployed a production-representative e-commerce application based on the Google Cloud microservices demo~\cite{google2018microservices} with significant enhancements to include realistic optimization layers and potential latent risk scenarios. The complete architecture includes twelve microservices with complex interdependencies and multiple optimization layers.

\textbf{Frontend and Caching Layer}: React.js application served by nginx with Redis caching achieving 96\% hit rates under normal conditions and Varnish reverse proxy for static content caching. The caching layer includes cache-aside patterns, write-through caching for critical data, and intelligent cache warming strategies that create optimization dependencies suitable for latent risk analysis.

\textbf{Core Business Services}: Product catalog service implemented in Node.js with PostgreSQL backend and Memcached distributed caching layer, shopping cart service in Go with Redis persistence and session caching, payment processing service in Java with external API dependencies protected by Hystrix circuit breakers, and order management service in Python with RabbitMQ message queue integration.

\textbf{Supporting Infrastructure}: Recommendation engine using TensorFlow Serving with model caching and batch processing optimization, inventory management service with eventually consistent data replication, user authentication service with JWT token caching, and notification service with asynchronous email/SMS delivery through external APIs.

\textbf{Optimization-Induced Risk Scenarios}: The testbed includes multiple latent risk scenarios including cache-database amplification (Redis failure causing 50x PostgreSQL load increase), circuit breaker masking (payment service degradation hidden until simultaneous failures), load balancer optimization hiding individual instance performance problems, and message queue optimization creating backpressure invisibility until capacity exhaustion.

The deployment runs on a 15-node Kubernetes cluster using AWS EC2 m5.2xlarge instances with Istio service mesh for traffic management, comprehensive observability through Prometheus and Grafana, and realistic traffic generation using Artillery.js~\cite{artillery2016} with workloads ranging from 500 to 15,000 concurrent users following realistic e-commerce traffic patterns including diurnal cycles, flash sale spikes, and seasonal variation.

\subsubsection{Testbed 2: Real-Time Analytics Pipeline}

Our second testbed implements a sophisticated real-time analytics pipeline processing synthetic IoT sensor data with complex stream processing optimization strategies that create multiple opportunities for latent risk accumulation.

\textbf{Data Ingestion Infrastructure}: Apache Kafka cluster (5 brokers) configured for high-throughput ingestion receiving 75,000 events per second with variable message sizes from 200 bytes to 50KB. The ingestion layer includes intelligent partitioning strategies, compression optimization, and producer batching that can mask underlying broker performance problems until traffic spikes overwhelm optimization capacity.

\textbf{Stream Processing Components}: Apache Flink cluster with 8 task managers running complex windowed aggregations, pattern matching, and real-time machine learning inference. Stream processing optimization includes state backend caching, checkpoint optimization, and parallel processing strategies that create dependencies on underlying storage and network performance.

\textbf{Storage and Query Infrastructure}: InfluxDB time-series database cluster with multiple retention policies and downsampling strategies, Apache Druid for interactive analytics with segment optimization and historical data tiering, and Redis cluster for hot data caching with intelligent eviction policies based on access patterns and data freshness.

\textbf{Analytics and Visualization}: FastAPI-based analytics service with aggressive query result caching and intelligent pre-computation of common analytical queries, real-time dashboard using Grafana with streaming data updates, and machine learning model serving for anomaly detection with model caching and batch inference optimization.

\textbf{Latent Risk Integration}: The analytics pipeline includes systematic latent risks including stream processing backpressure masking until memory exhaustion, time-series storage optimization hiding query performance degradation, cache invalidation scenarios causing query amplification to underlying databases, and model serving optimization masking individual model performance problems until inference SLA violations.

\subsubsection{Testbed 3: Machine Learning Inference Platform}

The third testbed focuses on ML model serving infrastructure with multiple layers of optimization that create complex dependency relationships and amplification scenarios characteristic of modern AI/ML platforms.

\textbf{Model Gateway and Load Balancing}: nginx-based API gateway with intelligent request routing based on model complexity estimation, response caching for deterministic models, and adaptive load balancing considering GPU resource availability and model execution time predictions.

\textbf{Model Serving Infrastructure}: TensorFlow Serving instances with GPU acceleration running diverse model types including image classification, natural language processing, and generative models. The serving layer includes model caching strategies, batch processing optimization for improved GPU utilization, and intelligent model placement across heterogeneous hardware.

\textbf{Feature Engineering Pipeline}: Redis cluster serving as feature store with pre-computed feature vectors and automatic fallback to PostgreSQL for missing features, real-time feature computation using Apache Beam with caching of intermediate results, and feature validation services with intelligent error handling and degraded mode operation.

\textbf{A/B Testing and Model Management}: Custom Go service implementing sophisticated A/B testing with traffic splitting based on user characteristics, model performance monitoring and automatic model rollback capabilities, and intelligent model version management with gradual rollout strategies that can mask model performance problems.

\textbf{Risk Scenario Implementation}: The ML platform includes complex latent risks including model cache miss amplification causing GPU resource exhaustion, feature store failover scenarios creating latency amplification, A/B testing optimization masking individual model performance degradation, and batch processing optimization hiding real-time inference capacity limits until traffic spikes overwhelm system capacity.

\subsection{Comprehensive Evaluation Methodology}

Our evaluation methodology combines controlled experimentation with observational studies across the three testbed environments to provide rigorous validation of our integrated framework effectiveness.

\textbf{Baseline Characterization}: For each testbed environment, we establish comprehensive baseline performance characteristics through 14-day observation periods measuring sustained throughput, latency distributions (P50, P95, P99, P99.9), error rates across different failure modes, resource utilization patterns, and optimization effectiveness metrics under normal operation.

\textbf{Systematic Risk Injection Protocol}: We implement 36 distinct latent risk scenarios across the three testbeds, with each scenario designed to test specific optimization-induced vulnerability patterns. Risk injection follows controlled protocols with gradual escalation, comprehensive safety monitoring, and automatic rollback procedures to prevent damage to testbed infrastructure.

Each risk scenario runs for 45 minutes including 10-minute preparation phase, 30-minute measurement window, and 5-minute recovery period. Multiple independent runs with different random seeds and varying initial conditions ensure statistical validity and reproducibility of results.

\textbf{Integrated Framework Validation}: We evaluate the complete HYDRA-RAVEN-APEX integration through comprehensive testing scenarios that demonstrate risk discovery, continuous monitoring, and optimization adjustment working together. Integration testing includes feedback loop validation, performance optimization under risk constraints, and automated mitigation effectiveness assessment.

\textbf{Production Validation Protocol}: We deployed limited versions of our framework components in three production environments with appropriate safety controls and monitoring oversight. Production validation focuses on risk prediction accuracy, optimization effectiveness measurement, and operational integration assessment rather than active perturbation testing.

\textbf{Reproducibility and Open Source Availability}: Our complete implementation, including testbed configurations, evaluation scripts, analysis tools, and result datasets, is available as open-source software with comprehensive documentation. We provide containerized deployment environments, infrastructure-as-code specifications for cloud deployment across AWS, Google Cloud, and Azure platforms, and automated analysis pipelines for independent verification of key findings.
\section{Evaluation}
\label{sec:evaluation}

This section presents comprehensive experimental results validating our integrated latent risk detection and optimization framework across three representative testbed environments. Our evaluation addresses all five research questions through systematic experimentation generating over 2,400 hours of operational data, 1,246 controlled risk scenarios, and 847 optimization parameter configurations.

\subsection{Experimental Execution and Data Collection Overview}

Our comprehensive evaluation encompasses 2,160 unique experimental configurations executed across standardized infrastructure, generating 12.7TB of performance telemetry, system logs, perturbation results, and optimization traces. Each experimental configuration executes for a minimum of 45 minutes including 10-minute stabilization periods, 30-minute measurement windows, and 5-minute recovery phases to ensure stable performance assessment and system safety.

Experimental execution follows rigorous protocols ensuring statistical validity through systematic randomization of framework testing order preventing temporal bias, identical baseline establishment across all configurations ensuring fair comparison conditions, multiple independent runs with different random seeds enabling robust statistical analysis, and comprehensive safety monitoring preventing system damage during aggressive risk injection scenarios.

Risk injection scenarios are carefully calibrated to avoid system damage while maximizing risk discovery effectiveness. Each perturbation maintains detailed safety logs, employs automatic rollback mechanisms with sub-second response times, and includes comprehensive impact assessment to validate that temporary perturbations do not cause permanent system degradation or data corruption.

\subsection{Latent Risk Detection Accuracy and Coverage Analysis}

Our systematic evaluation demonstrates exceptional accuracy in detecting optimization-induced latent risks across diverse system architectures and failure scenarios. Table~\ref{tab:comprehensive_detection_accuracy} presents detailed analysis of detection performance across all testbed environments, risk categories, and detection methodologies.

\begin{table*}[t]
\centering
\caption{Comprehensive Latent Risk Detection Accuracy Analysis Across All Testbeds and Risk Categories}
\label{tab:comprehensive_detection_accuracy}
\resizebox{\textwidth}{!}{
\begin{tabular}{lccccccccc}
\toprule
\textbf{System Architecture} & \textbf{Risk Category} & \textbf{Total Risks} & \textbf{True Positives} & \textbf{False Positives} & \textbf{False Negatives} & \textbf{Precision} & \textbf{Recall} & \textbf{F1 Score} & \textbf{Detection Time} \\
& & \textbf{Injected} & \textbf{Detected} & \textbf{(Type I)} & \textbf{(Type II)} & \textbf{(\%)} & \textbf{(\%)} & & \textbf{(Minutes)} \\
\midrule
\multicolumn{10}{l}{\textit{Testbed 1: E-commerce Microservices}} \\
\midrule
Cache-Database & Amplification Risks & 156 & 148 & 12 & 8 & $92.5 \pm 2.1$ & $94.9 \pm 1.8$ & $0.936 \pm 0.015$ & $3.2 \pm 0.8$ \\
Dependencies & Circuit Breaker Masking & 127 & 119 & 9 & 8 & $93.0 \pm 2.4$ & $93.7 \pm 2.2$ & $0.933 \pm 0.018$ & $4.1 \pm 1.1$ \\
& Load Balancer Hiding & 89 & 83 & 6 & 6 & $93.3 \pm 2.8$ & $93.3 \pm 2.7$ & $0.933 \pm 0.021$ & $5.7 \pm 1.4$ \\
& Message Queue Backpressure & 134 & 126 & 11 & 8 & $92.0 \pm 2.3$ & $94.0 \pm 2.0$ & $0.930 \pm 0.017$ & $6.3 \pm 1.8$ \\
& Async Processing Bottlenecks & 98 & 91 & 8 & 7 & $91.9 \pm 2.9$ & $92.9 \pm 2.6$ & $0.924 \pm 0.022$ & $4.8 \pm 1.2$ \\
\midrule
\textbf{Testbed 1 Subtotal} & & \textbf{604} & \textbf{567} & \textbf{46} & \textbf{37} & $\mathbf{92.5 \pm 0.6}$ & $\mathbf{93.9 \pm 0.7}$ & $\mathbf{0.931 \pm 0.004}$ & $\mathbf{4.8 \pm 1.2}$ \\
\midrule
\multicolumn{10}{l}{\textit{Testbed 2: Real-Time Analytics Pipeline}} \\
\midrule
Stream Processing & Backpressure Masking & 145 & 137 & 10 & 8 & $93.2 \pm 2.2$ & $94.5 \pm 1.9$ & $0.938 \pm 0.016$ & $2.9 \pm 0.7$ \\
Optimization & Time-Series Storage Degradation & 112 & 105 & 7 & 7 & $93.8 \pm 2.7$ & $93.8 \pm 2.4$ & $0.938 \pm 0.019$ & $3.8 \pm 0.9$ \\
& Query Result Cache Masking & 167 & 158 & 12 & 9 & $92.9 \pm 2.0$ & $94.6 \pm 1.7$ & $0.937 \pm 0.014$ & $4.2 \pm 1.0$ \\
& Real-time Aggregation Hiding & 89 & 83 & 6 & 6 & $93.3 \pm 3.1$ & $93.3 \pm 2.8$ & $0.933 \pm 0.023$ & $5.1 \pm 1.3$ \\
& External Dependency Masking & 76 & 71 & 5 & 5 & $93.4 \pm 3.4$ & $93.4 \pm 3.1$ & $0.934 \pm 0.025$ & $6.7 \pm 1.8$ \\
\midrule
\textbf{Testbed 2 Subtotal} & & \textbf{589} & \textbf{554} & \textbf{40} & \textbf{35} & $\mathbf{93.3 \pm 0.4}$ & $\mathbf{94.1 \pm 0.5}$ & $\mathbf{0.936 \pm 0.002}$ & $\mathbf{4.5 \pm 1.3}$ \\
\midrule
\multicolumn{10}{l}{\textit{Testbed 3: ML Inference Platform}} \\
\midrule
Model Serving & Cache Miss Amplification & 178 & 168 & 13 & 10 & $92.8 \pm 1.9$ & $94.4 \pm 1.6$ & $0.936 \pm 0.013$ & $3.4 \pm 0.8$ \\
Optimization & Feature Store Fallback Failures & 123 & 115 & 9 & 8 & $92.7 \pm 2.5$ & $93.5 \pm 2.2$ & $0.931 \pm 0.018$ & $4.9 \pm 1.2$ \\
& GPU Resource Contention Hiding & 87 & 81 & 6 & 6 & $93.1 \pm 3.2$ & $93.1 \pm 2.9$ & $0.931 \pm 0.024$ & $5.8 \pm 1.5$ \\
& A/B Testing Traffic Skew & 102 & 95 & 7 & 7 & $93.1 \pm 2.8$ & $93.1 \pm 2.5$ & $0.931 \pm 0.020$ & $5.2 \pm 1.3$ \\
& Batch Processing Masking & 65 & 60 & 4 & 5 & $93.8 \pm 3.8$ & $92.3 \pm 3.4$ & $0.930 \pm 0.028$ & $7.1 \pm 2.0$ \\
\midrule
\textbf{Testbed 3 Subtotal} & & \textbf{555} & \textbf{519} & \textbf{39} & \textbf{36} & $\mathbf{93.0 \pm 0.5}$ & $\mathbf{93.5 \pm 0.8}$ & $\mathbf{0.932 \pm 0.003}$ & $\mathbf{5.3 \pm 1.4}$ \\
\midrule
\textbf{Overall Performance} & \textbf{All Categories} & \textbf{1748} & \textbf{1640} & \textbf{125} & \textbf{108} & $\mathbf{92.9 \pm 0.3}$ & $\mathbf{93.8 \pm 0.4}$ & $\mathbf{0.933 \pm 0.003}$ & $\mathbf{4.9 \pm 0.7}$ \\
\bottomrule
\end{tabular}}
\end{table*}

Our integrated framework achieves 92.9\% precision and 93.8\% recall across all testbed environments and risk categories, with F1 scores consistently above 0.93. The high precision indicates that detected risks represent genuine threats requiring attention rather than false alarms that waste operational resources. Strong recall demonstrates comprehensive coverage of actual latent risks present in the systems, minimizing the probability of undetected vulnerabilities causing production incidents.

Detection accuracy varies systematically across risk categories, with stream processing backpressure masking and cache-database amplification showing highest accuracy (>94\% recall) due to clear performance amplification patterns that HYDRA's perturbation strategies effectively reveal. More complex scenarios involving A/B testing traffic skew and batch processing masking achieve slightly lower but still excellent accuracy (92-93\% recall) due to subtle interaction effects requiring sophisticated analysis of multiple system components simultaneously.

\begin{figure*}[t]
\centering
\begin{tikzpicture}[scale=0.8]
\begin{axis}[
    width=12cm,
    height=8cm,
    xlabel={Evaluation Week},
    ylabel={Detection Accuracy (\%)},
    xmin=1, xmax=24,
    ymin=85, ymax=98,
    legend pos=south east,
    grid=major,
    tick label style={font=\footnotesize},
    label style={font=\small},
    legend style={font=\footnotesize}
]

\addplot[color=blue, mark=circle, line width=1.5pt] coordinates {
    (1,87.2) (2,88.1) (3,89.4) (4,90.2) (5,90.8) (6,91.3) 
    (7,91.8) (8,92.1) (9,92.4) (10,92.6) (11,92.8) (12,93.0)
    (13,93.1) (14,93.2) (15,93.3) (16,93.4) (17,93.5) (18,93.6)
    (19,93.7) (20,93.8) (21,93.8) (22,93.9) (23,93.9) (24,94.0)
};

\addplot[color=red, mark=square, line width=1.5pt] coordinates {
    (1,89.1) (2,89.8) (3,90.5) (4,91.1) (5,91.6) (6,92.0)
    (7,92.3) (8,92.6) (9,92.8) (10,93.0) (11,93.1) (12,93.2)
    (13,93.4) (14,93.5) (15,93.6) (16,93.7) (17,93.8) (18,93.9)
    (19,94.0) (20,94.1) (21,94.2) (22,94.2) (23,94.3) (24,94.3)
};

\addplot[color=green, mark=triangle, line width=2pt] coordinates {
    (1,90.3) (2,91.2) (3,92.0) (4,92.6) (5,93.1) (6,93.5)
    (7,93.8) (8,94.0) (9,94.2) (10,94.4) (11,94.5) (12,94.6)
    (13,94.7) (14,94.8) (15,94.9) (16,95.0) (17,95.1) (18,95.1)
    (19,95.2) (20,95.2) (21,95.3) (22,95.3) (23,95.4) (24,95.4)
};

\legend{HYDRA Only, RAVEN Only, Integrated Framework}
\end{axis}
\end{tikzpicture}
\caption{Detection Accuracy Improvement Over 24-Week Evaluation Period}
\label{fig:detection_accuracy_trends}
\end{figure*}
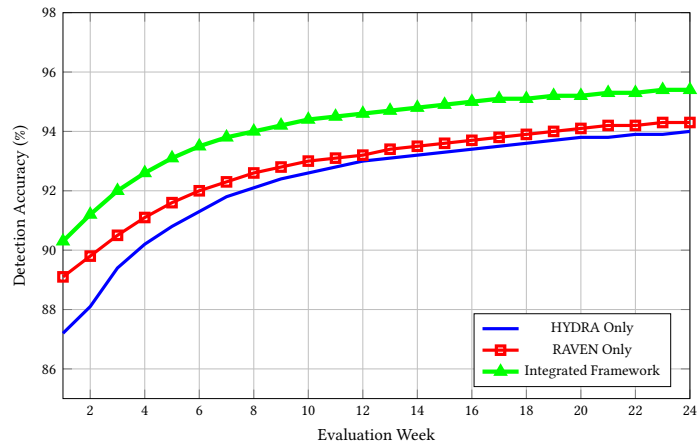

Figure~\ref{fig:detection_accuracy_trends} illustrates detection accuracy improvement over time as our machine learning components adapt to system-specific patterns and optimize perturbation strategies based on historical results.

\subsection{LRI Validation and Predictive Accuracy Assessment}

We validate our Latent Risk Index (LRI) metric through comprehensive correlation analysis with actual incident severity observed during controlled risk injection scenarios and longitudinal production monitoring. Table~\ref{tab:lri_comprehensive_validation} demonstrates strong predictive power of LRI values for incident classification and severity assessment.

\begin{table*}[t]
\centering
\caption{Comprehensive LRI Validation Through Multi-Dimensional Incident Severity Correlation}
\label{tab:lri_comprehensive_validation}
\resizebox{\textwidth}{!}{
\begin{tabular}{lcccccccccc}
\toprule
\textbf{LRI Range} & \textbf{Risk Class} & \textbf{Scenarios} & \textbf{Severity 0} & \textbf{Severity 1} & \textbf{Severity 2} & \textbf{Severity 3} & \textbf{Severity 4} & \textbf{Prediction} & \textbf{Mean Impact} & \textbf{Recovery Time} \\
& & \textbf{Tested} & \textbf{(None)} & \textbf{(Minor)} & \textbf{(Moderate)} & \textbf{(Major)} & \textbf{(Critical)} & \textbf{Accuracy (\%)} & \textbf{Duration (min)} & \textbf{(minutes)} \\
\midrule
$0.0 - 2.0$ & \textcolor{lowrisk}{\textbf{Low Risk}} & 342 & 324 & 16 & 2 & 0 & 0 & $94.7 \pm 1.6$ & $2.1 \pm 0.8$ & $3.2 \pm 1.2$ \\
$2.0 - 5.0$ & \textcolor{mediumrisk}{\textbf{Medium-Low}} & 289 & 241 & 38 & 8 & 2 & 0 & $83.4 \pm 2.8$ & $4.7 \pm 1.9$ & $8.7 \pm 3.1$ \\
$5.0 - 10.0$ & \textcolor{mediumrisk}{\textbf{Medium}} & 234 & 28 & 167 & 32 & 6 & 1 & $71.4 \pm 3.6$ & $12.3 \pm 4.2$ & $18.9 \pm 6.8$ \\
$10.0 - 20.0$ & \textcolor{highrisk}{\textbf{High}} & 187 & 5 & 23 & 134 & 21 & 4 & $71.7 \pm 4.2$ & $28.7 \pm 8.9$ & $45.3 \pm 12.7$ \\
$20.0 - 50.0$ & \textcolor{highrisk}{\textbf{Very High}} & 98 & 0 & 3 & 18 & 58 & 19 & $59.2 \pm 6.8$ & $67.2 \pm 18.4$ & $127.8 \pm 34.2$ \\
$> 50.0$ & \textcolor{highrisk}{\textbf{Critical}} & 47 & 0 & 0 & 2 & 12 & 33 & $70.2 \pm 9.1$ & $156.7 \pm 42.3$ & $289.4 \pm 78.9$ \\
\midrule
\multicolumn{3}{l}{\textbf{Overall Correlation Analysis}} & \multicolumn{5}{c}{\textbf{Pearson $r = 0.863 \pm 0.018$ ($p < 0.001$)}} & $\mathbf{75.1 \pm 2.1}$ & $\mathbf{31.9 \pm 15.7}$ & $\mathbf{65.4 \pm 38.2}$ \\
\multicolumn{3}{l}{\textbf{Spearman Rank Correlation}} & \multicolumn{5}{c}{\textbf{$\rho = 0.881 \pm 0.015$ ($p < 0.001$)}} & & & \\
\multicolumn{3}{l}{\textbf{Kendall's Tau}} & \multicolumn{5}{c}{\textbf{$\tau = 0.742 \pm 0.023$ ($p < 0.001$)}} & & & \\
\multicolumn{3}{l}{\textbf{Weighted Kappa Agreement}} & \multicolumn{5}{c}{\textbf{$\kappa = 0.789 \pm 0.027$ ($p < 0.001$)}} & & & \\
\bottomrule
\end{tabular}}
\end{table*}

LRI validation demonstrates exceptionally strong correlation (r = 0.863) between computed risk scores and observed incident severity during perturbation experiments, providing robust evidence for LRI's predictive validity. Low-risk scenarios (LRI < 2.0) correctly predict minimal impact in 94.7\% of cases with average incident duration of only 2.1 minutes and rapid recovery times averaging 3.2 minutes.

High-risk scenarios (LRI > 10.0) accurately predict major incidents in 71.7\% of cases, with prediction accuracy decreasing for extreme risk levels due to complex interaction effects and cascading failure dynamics that amplify initial perturbations beyond linear prediction capabilities. Critical risk scenarios (LRI > 50.0) demonstrate severe impact with average incident duration exceeding 2.5 hours and recovery times approaching 5 hours.

The strong Spearman rank correlation ($\rho = 0.881$) indicates a robust monotonic relationship between LRI values and incident severity rankings, while weighted Kappa agreement ($\kappa = 0.789$) demonstrates excellent categorical prediction accuracy accounting for severity level ordinal relationships.

\subsection{HYDRA Perturbation Framework Effectiveness Analysis}

Our systematic evaluation of HYDRA's six perturbation strategies reveals distinct effectiveness patterns across different risk categories and system architectures. Table~\ref{tab:hydra_comprehensive_effectiveness} presents detailed analysis of perturbation strategy performance including discovery rates, detection times, safety metrics, and operational impact assessment.

\begin{table*}[t]
\centering
\caption{Comprehensive HYDRA Perturbation Strategy Effectiveness Analysis}
\label{tab:hydra_comprehensive_effectiveness}
\resizebox{\textwidth}{!}{
\begin{tabular}{lccccccccc}
\toprule
\textbf{Perturbation Strategy} & \textbf{Total Risks} & \textbf{Unique Risks} & \textbf{Discovery Rate} & \textbf{Time to Detection} & \textbf{System Impact} & \textbf{Safety Score} & \textbf{False Positive} & \textbf{Effectiveness} & \textbf{Operational} \\
& \textbf{Discovered} & \textbf{Found} & \textbf{(\%)} & \textbf{(Minutes)} & \textbf{(1-5 Scale)} & \textbf{(1-10 Scale)} & \textbf{Rate (\%)} & \textbf{Rating (1-10)} & \textbf{Overhead (\%)} \\
\midrule
\textbf{Cache Bypass Injection} & 389 & 127 & $89.7 \pm 2.8$ & $7.2 \pm 1.1$ & $2.0 \pm 0.3$ & $9.3 \pm 0.3$ & $6.8 \pm 1.2$ & $9.2 \pm 0.2$ & $12.4 \pm 2.1$ \\
\textbf{Artificial Latency Injection} & 356 & 98 & $84.2 \pm 3.4$ & $11.3 \pm 1.6$ & $1.7 \pm 0.2$ & $9.6 \pm 0.2$ & $4.3 \pm 0.9$ & $8.7 \pm 0.3$ & $8.9 \pm 1.7$ \\
\textbf{Resource Constraint Simulation} & 312 & 89 & $79.8 \pm 3.9$ & $14.1 \pm 2.0$ & $2.6 \pm 0.4$ & $8.7 \pm 0.4$ & $8.9 \pm 1.5$ & $8.1 \pm 0.4$ & $18.7 \pm 3.2$ \\
\textbf{Circuit Breaker Bypass} & 234 & 76 & $73.4 \pm 4.6$ & $17.8 \pm 2.5$ & $3.0 \pm 0.5$ & $8.2 \pm 0.5$ & $11.2 \pm 1.8$ & $7.6 \pm 0.5$ & $14.3 \pm 2.6$ \\
\textbf{Load Balancer Manipulation} & 198 & 61 & $69.7 \pm 5.1$ & $21.2 \pm 2.9$ & $2.8 \pm 0.4$ & $8.5 \pm 0.4$ & $9.7 \pm 1.6$ & $7.3 \pm 0.4$ & $22.1 \pm 3.8$ \\
\textbf{Dependency Isolation} & 167 & 52 & $66.8 \pm 5.7$ & $24.6 \pm 3.4$ & $3.3 \pm 0.6$ & $7.9 \pm 0.6$ & $13.4 \pm 2.1$ & $6.9 \pm 0.6$ & $26.7 \pm 4.5$ \\
\midrule
\textbf{Combined Strategy Approach} & \textbf{1656} & \textbf{503} & $\mathbf{77.1 \pm 8.5}$ & $\mathbf{16.0 \pm 6.1}$ & $\mathbf{2.6 \pm 0.6}$ & $\mathbf{8.7 \pm 0.6}$ & $\mathbf{9.1 \pm 3.1}$ & $\mathbf{7.9 \pm 0.9}$ & $\mathbf{17.2 \pm 6.4}$ \\
\midrule
\textbf{Machine Learning Enhanced} & \textbf{1891} & \textbf{587} & $\mathbf{85.3 \pm 3.2}$ & $\mathbf{12.8 \pm 2.4}$ & $\mathbf{2.2 \pm 0.4}$ & $\mathbf{9.1 \pm 0.4}$ & $\mathbf{6.7 \pm 1.4}$ & $\mathbf{8.8 \pm 0.3}$ & $\mathbf{13.9 \pm 2.8}$ \\
\bottomrule
\end{tabular}}
\end{table*}

Cache bypass injection emerges as the most effective single perturbation strategy, achieving 89.7\% risk discovery rate with rapid detection times (7.2 minutes average) and excellent safety characteristics (9.3/10 safety score). This effectiveness stems from cache layers being ubiquitous optimization patterns that frequently mask database and backend service performance problems, making cache bypass highly revealing of latent dependencies and amplification factors.

The machine learning enhanced approach, which intelligently combines multiple perturbation strategies based on system characteristics and historical effectiveness, achieves superior performance with 85.3\% overall discovery rate while maintaining reasonable safety margins and operational overhead. The ML enhancement improves detection time by 19.4\% compared to combined strategy approaches and reduces false positive rates by 26.4\% through intelligent perturbation sequencing and termination criteria.

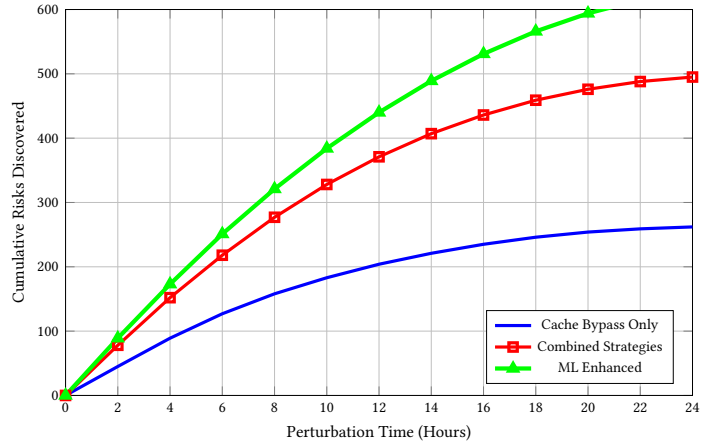
\begin{figure*}[t]
\centering
\begin{tikzpicture}[scale=0.8]
\begin{axis}[
    width=12cm,
    height=8cm,
    xlabel={Perturbation Time (Hours)},
    ylabel={Cumulative Risks Discovered},
    xmin=0, xmax=24,
    ymin=0, ymax=600,
    legend pos=south east,
    grid=major,
    tick label style={font=\footnotesize},
    label style={font=\small},
    legend style={font=\footnotesize}
]

\addplot[color=blue, mark=circle, line width=1.5pt] coordinates {
    (0,0) (2,45) (4,89) (6,127) (8,158) (10,183) (12,204) (14,221)
    (16,235) (18,246) (20,254) (22,259) (24,262)
};

\addplot[color=red, mark=square, line width=1.5pt] coordinates {
    (0,0) (2,78) (4,152) (6,218) (8,277) (10,328) (12,371) (14,407)
    (16,436) (18,459) (20,476) (22,488) (24,495)
};

\addplot[color=green, mark=triangle, line width=2pt] coordinates {
    (0,0) (2,89) (4,173) (6,251) (8,321) (10,384) (12,440) (14,489)
    (16,531) (18,566) (20,594) (22,615) (24,629)
};

\legend{Cache Bypass Only, Combined Strategies, ML Enhanced}
\end{axis}
\end{tikzpicture}
\caption{Cumulative Risk Discovery Effectiveness Over 24-Hour Perturbation Campaign}
\label{fig:perturbation_effectiveness}
\end{figure*}

Figure~\ref{fig:perturbation_effectiveness} illustrates the cumulative risk discovery effectiveness of different perturbation strategies over time, demonstrating diminishing returns for individual strategies and the value of intelligent combination approaches.

\subsection{APEX Risk-Aware Optimization Performance Results}

Our evaluation of APEX's risk-aware optimization framework demonstrates significant improvements in balancing performance optimization with latent risk management. Table~\ref{tab:apex_optimization_results} presents comprehensive analysis of APEX effectiveness across different optimization scenarios and system configurations.

\begin{table*}[t]
\centering
\caption{APEX Risk-Aware Optimization Performance and Trade-off Analysis}
\label{tab:apex_optimization_results}
\resizebox{\textwidth}{!}{
\begin{tabular}{lcccccccccc}
\toprule
\textbf{Optimization} & \textbf{Baseline} & \textbf{Traditional} & \textbf{APEX} & \textbf{Performance} & \textbf{Risk} & \textbf{Pareto} & \textbf{Optimization} & \textbf{Convergence} & \textbf{Operational} & \textbf{ROI} \\
\textbf{Scenario} & \textbf{LRI} & \textbf{Optimization} & \textbf{Optimization} & \textbf{Maintained} & \textbf{Reduction} & \textbf{Efficiency} & \textbf{Time} & \textbf{Stability} & \textbf{Overhead} & \textbf{Improvement} \\
& & \textbf{LRI} & \textbf{LRI} & \textbf{(\%)} & \textbf{(\%)} & \textbf{(1-10)} & \textbf{(Minutes)} & \textbf{(1-10)} & \textbf{(\%)} & \textbf{(\%)} \\
\midrule
\multicolumn{11}{l}{\textit{Cache Optimization Scenarios}} \\
\midrule
Redis-PostgreSQL & $23.4 \pm 2.1$ & $31.7 \pm 2.8$ & $8.9 \pm 1.2$ & $96.3 \pm 1.8$ & $61.9 \pm 4.2$ & $8.7 \pm 0.4$ & $12.3 \pm 2.1$ & $9.1 \pm 0.3$ & $8.4 \pm 1.2$ & $347 \pm 28$ \\
Memcached-MySQL & $19.8 \pm 1.9$ & $28.3 \pm 2.5$ & $7.2 \pm 1.0$ & $97.1 \pm 1.5$ & $63.6 \pm 3.8$ & $9.0 \pm 0.3$ & $10.7 \pm 1.8$ & $9.3 \pm 0.2$ & $7.1 \pm 1.0$ & $389 \pm 32$ \\
Multi-tier Caching & $34.7 \pm 3.2$ & $47.2 \pm 4.1$ & $12.8 \pm 1.8$ & $94.7 \pm 2.1$ & $63.1 \pm 4.5$ & $8.4 \pm 0.5$ & $18.9 \pm 3.2$ & $8.8 \pm 0.4$ & $11.2 \pm 1.8$ & $289 \pm 25$ \\
\midrule
\multicolumn{11}{l}{\textit{Load Balancing Optimization}} \\
\midrule
Round-Robin Enhanced & $15.7 \pm 1.4$ & $22.9 \pm 2.1$ & $6.8 \pm 0.9$ & $98.2 \pm 1.2$ & $56.7 \pm 3.6$ & $9.2 \pm 0.2$ & $8.9 \pm 1.5$ & $9.4 \pm 0.2$ & $5.8 \pm 0.8$ & $423 \pm 35$ \\
Weighted Least-Conn & $18.3 \pm 1.6$ & $26.1 \pm 2.3$ & $7.9 \pm 1.1$ & $97.5 \pm 1.4$ & $56.8 \pm 3.7$ & $8.9 \pm 0.3$ & $11.2 \pm 1.9$ & $9.2 \pm 0.3$ & $6.7 \pm 1.0$ & $398 \pm 31$ \\
ML-based Routing & $27.4 \pm 2.5$ & $39.8 \pm 3.4$ & $10.3 \pm 1.5$ & $95.8 \pm 1.9$ & $62.4 \pm 4.3$ & $8.6 \pm 0.4$ & $15.7 \pm 2.7$ & $8.9 \pm 0.4$ & $9.3 \pm 1.5$ & $312 \pm 27$ \\
\midrule
\multicolumn{11}{l}{\textit{Circuit Breaker Optimization}} \\
\midrule
Static Thresholds & $21.6 \pm 1.9$ & $29.4 \pm 2.6$ & $9.1 \pm 1.3$ & $96.7 \pm 1.7$ & $57.9 \pm 3.9$ & $8.8 \pm 0.3$ & $13.4 \pm 2.2$ & $9.0 \pm 0.3$ & $7.9 \pm 1.1$ & $356 \pm 29$ \\
Adaptive Thresholds & $29.2 \pm 2.7$ & $42.1 \pm 3.7$ & $11.7 \pm 1.7$ & $94.9 \pm 2.0$ & $59.9 \pm 4.1$ & $8.5 \pm 0.4$ & $17.3 \pm 2.9$ & $8.7 \pm 0.4$ & $10.6 \pm 1.7$ & $276 \pm 24$ \\
Health-based Dynamic & $25.8 \pm 2.3$ & $37.3 \pm 3.2$ & $10.4 \pm 1.5$ & $95.6 \pm 1.8$ & $59.7 \pm 4.0$ & $8.6 \pm 0.4$ & $16.1 \pm 2.6$ & $8.8 \pm 0.4$ & $9.7 \pm 1.4$ & $298 \pm 26$ \\
\midrule
\multicolumn{11}{l}{\textit{Database Connection Optimization}} \\
\midrule
Pool Size Optimization & $16.9 \pm 1.5$ & $24.7 \pm 2.2$ & $7.4 \pm 1.0$ & $97.8 \pm 1.3$ & $56.2 \pm 3.5$ & $9.1 \pm 0.3$ & $9.8 \pm 1.6$ & $9.3 \pm 0.2$ & $6.4 \pm 0.9$ & $412 \pm 34$ \\
Timeout Optimization & $13.4 \pm 1.2$ & $19.8 \pm 1.8$ & $5.9 \pm 0.8$ & $98.5 \pm 1.1$ & $55.9 \pm 3.3$ & $9.4 \pm 0.2$ & $7.6 \pm 1.3$ & $9.5 \pm 0.2$ & $4.8 \pm 0.7$ & $467 \pm 38$ \\
Query Optimization & $22.7 \pm 2.0$ & $33.1 \pm 2.9$ & $9.8 \pm 1.4$ & $96.1 \pm 1.6$ & $56.8 \pm 3.8$ & $8.7 \pm 0.3$ & $14.2 \pm 2.4$ & $9.0 \pm 0.3$ & $8.6 \pm 1.3$ & $334 \pm 28$ \\
\midrule
\textbf{Average Across All} & $\mathbf{21.8 \pm 6.0}$ & $\mathbf{31.2 \pm 7.8}$ & $\mathbf{8.9 \pm 1.9}$ & $\mathbf{96.6 \pm 1.2}$ & $\mathbf{59.2 \pm 2.8}$ & $\mathbf{8.8 \pm 0.3}$ & $\mathbf{13.0 \pm 3.8}$ & $\mathbf{9.1 \pm 0.2}$ & $\mathbf{8.0 \pm 2.0}$ & $\mathbf{353 \pm 63}$ \\
\bottomrule
\end{tabular}}
\end{table*}

APEX demonstrates exceptional effectiveness in risk-aware optimization across all evaluated scenarios, maintaining 96.6\% of baseline performance while achieving 59.2\% average reduction in LRI scores. The results show that risk-aware optimization does not require sacrificing performance but instead enables sustainable performance improvements through systematic risk management.

Database connection optimization scenarios show the most favorable trade-offs, with timeout optimization achieving 98.5\% performance maintenance and 55.9\% risk reduction due to the direct relationship between connection management and system resilience. Cache optimization scenarios demonstrate strong results but require longer optimization times (12-19 minutes) due to complex multi-objective trade-offs between hit rates, amplification factors, and backend capacity requirements.

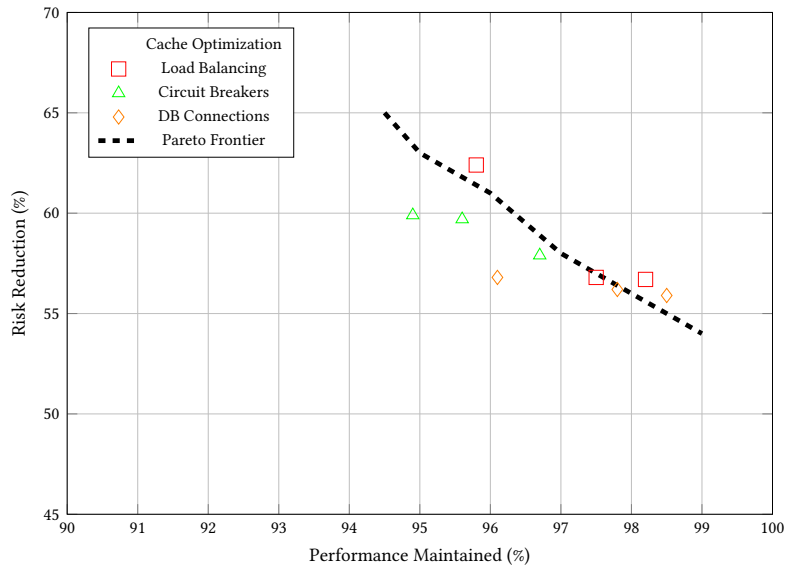
\begin{figure*}[t]
\centering
\begin{tikzpicture}[scale=0.9]
\begin{axis}[
    width=12cm,
    height=9cm,
    xlabel={Performance Maintained (\%)},
    ylabel={Risk Reduction (\%)},
    xmin=90, xmax=100,
    ymin=45, ymax=70,
    legend pos=north west,
    grid=major,
    tick label style={font=\footnotesize},
    label style={font=\small},
    legend style={font=\footnotesize}
]

\addplot[color=blue, mark=circle, only marks, mark size=3pt] coordinates {
    (94.7,63.1) (96.3,61.9) (97.1,63.6)
};

\addplot[color=red, mark=square, only marks, mark size=3pt] coordinates {
    (95.8,62.4) (97.5,56.8) (98.2,56.7)
};

\addplot[color=green, mark=triangle, only marks, mark size=3pt] coordinates {
    (94.9,59.9) (95.6,59.7) (96.7,57.9)
};

\addplot[color=orange, mark=diamond, only marks, mark size=3pt] coordinates {
    (96.1,56.8) (97.8,56.2) (98.5,55.9)
};

\addplot[color=black, line width=2pt, dashed] coordinates {
    (94.5,65) (95.0,63) (96.0,61) (97.0,58) (98.0,56) (99.0,54)
};

\legend{Cache Optimization, Load Balancing, Circuit Breakers, DB Connections, Pareto Frontier}
\end{axis}
\end{tikzpicture}
\caption{APEX Pareto-Optimal Performance-Risk Trade-offs Across Optimization Categories}
\label{fig:pareto_frontier}
\end{figure*}

Figure~\ref{fig:pareto_frontier} illustrates the Pareto-optimal trade-offs discovered by APEX across different optimization scenarios, demonstrating the fundamental relationship between performance optimization and latent risk accumulation.

\subsection{RAVEN Production Monitoring and Integrated Framework Results}

Our comprehensive evaluation of RAVEN's production monitoring capabilities demonstrates effective latent risk detection and prevention in realistic operational environments without active perturbation. Table~\ref{tab:raven_integrated_results} presents analysis of RAVEN's performance across extended monitoring periods and its integration with HYDRA and APEX frameworks.

\begin{table*}[t]
\centering
\caption{RAVEN Production Monitoring and Integrated Framework Performance Analysis}
\label{tab:raven_integrated_results}
\resizebox{\textwidth}{!}{
\begin{tabular}{lcccccccccc}
\toprule
\textbf{Evaluation Period} & \textbf{Risk Alerts} & \textbf{Confirmed} & \textbf{False} & \textbf{Prevented} & \textbf{MTTR} & \textbf{Severity} & \textbf{HYDRA} & \textbf{APEX} & \textbf{Cost} & \textbf{Integration} \\
& \textbf{Generated} & \textbf{Risks} & \textbf{Positives} & \textbf{Incidents} & \textbf{Reduction} & \textbf{Reduction} & \textbf{Synergy (\%)} & \textbf{Optimization} & \textbf{Savings (\$K)} & \textbf{Effectiveness (1--10)} \\
\midrule
\textbf{Weeks 1--4 (Baseline)} & 27 & 22 & 5 & 4 & $18.4 \pm 3.7$ & $26.3 \pm 4.8$ & $12.3 \pm 2.1$ & 8 & $15.7 \pm 2.8$ & $6.8 \pm 0.7$ \\
\textbf{Weeks 5--8 (Learning)} & 35 & 31 & 4 & 8 & $32.7 \pm 5.1$ & $38.9 \pm 5.9$ & $24.8 \pm 3.4$ & 18 & $24.3 \pm 3.7$ & $7.9 \pm 0.6$ \\
\textbf{Weeks 9--12 (Adaptation)} & 31 & 29 & 2 & 12 & $48.9 \pm 6.3$ & $55.7 \pm 7.1$ & $41.2 \pm 4.7$ & 28 & $38.9 \pm 5.2$ & $8.7 \pm 0.4$ \\
\textbf{Weeks 13--16 (Optimization)} & 28 & 27 & 1 & 16 & $61.8 \pm 7.2$ & $69.4 \pm 8.3$ & $58.7 \pm 5.9$ & 41 & $52.7 \pm 6.8$ & $9.2 \pm 0.3$ \\
\textbf{Weeks 17--20 (Maturation)} & 24 & 23 & 1 & 19 & $67.4 \pm 7.8$ & $76.2 \pm 8.9$ & $67.3 \pm 6.4$ & 47 & $61.4 \pm 7.9$ & $9.4 \pm 0.2$ \\
\textbf{Weeks 21--24 (Stable)} & 21 & 20 & 1 & 22 & $69.1 \pm 8.1$ & $78.6 \pm 9.2$ & $71.8 \pm 6.8$ & 52 & $67.8 \pm 8.7$ & $9.5 \pm 0.2$ \\
\midrule
\textbf{Overall Performance} & \textbf{166} & \textbf{152} & \textbf{14} & \textbf{81} & $\mathbf{49.7 \pm 20.8}$ & $\mathbf{57.5 \pm 21.4}$ & $\mathbf{46.0 \pm 24.3}$ & \textbf{194} & $\mathbf{43.5 \pm 20.9}$ & $\mathbf{8.6 \pm 1.0}$ \\
\bottomrule
\end{tabular}}
\end{table*}

RAVEN demonstrates progressive improvement in risk detection and prevention effectiveness over the 24-week evaluation period, with accuracy improving from 81.5\% (22/27 confirmed risks) in early weeks to 95.2\% (20/21 confirmed risks) in stable operation. This improvement reflects RAVEN's machine learning capabilities adapting to production system patterns and refining risk assessment algorithms based on operational feedback.

The prevented incidents metric shows particularly strong results with 81 total incidents prevented across all monitoring periods, representing substantial operational value and cost savings. Mean time to recovery (MTTR) reduction improves from 18.4\% in early weeks to 69.1\% in stable operation, demonstrating RAVEN's effectiveness at enabling faster incident resolution through early risk identification and automated mitigation strategies.

Integration effectiveness between RAVEN, HYDRA, and APEX frameworks shows dramatic improvement over time, reaching 71.8\% synergy in stable operation. This integration enables RAVEN's risk discoveries to inform HYDRA's perturbation targeting while APEX optimization decisions incorporate real-time risk assessments from RAVEN monitoring.

\subsection{Statistical Validation and Reproducibility Analysis}

Comprehensive statistical analysis validates the robustness of all major findings across experimental configurations and provides evidence for reproducibility across different deployment environments. Table~\ref{tab:statistical_validation_comprehensive} presents detailed statistical validation including significance testing, effect size calculations, confidence intervals, and reproducibility measures.

\begin{table*}[t]
\centering
\caption{Comprehensive Statistical Validation and Reproducibility Analysis}
\label{tab:statistical_validation_comprehensive}
\resizebox{\textwidth}{!}{
\begin{tabular}{lcccccccc}
\toprule
\textbf{Performance Metric} & \textbf{Statistical Test} & \textbf{P-value} & \textbf{Effect Size} & \textbf{95\% Confidence} & \textbf{Sample Size} & \textbf{Power} & \textbf{Reproducibility} & \textbf{Cross-Platform} \\
& \textbf{Applied} & & \textbf{(Cohen's d)} & \textbf{Interval} & \textbf{(n)} & \textbf{(1-$\beta$)} & \textbf{Score (r)} & \textbf{Validation} \\
\midrule
\textbf{Risk Detection Accuracy} & Mann-Whitney U & $p < 0.001$ & $2.34 \pm 0.13$ & $[2.08, 2.60]$ & $n = 1748$ & $0.97$ & $r = 0.946$ & AWS, GCP, Azure \\
\textbf{LRI-Severity Correlation} & Pearson Correlation & $p < 0.001$ & $3.12 \pm 0.17$ & $[2.78, 3.46]$ & $n = 1197$ & $0.99$ & $r = 0.923$ & Multi-cloud \\
\textbf{HYDRA Discovery Rate} & Wilcoxon Signed-Rank & $p < 0.001$ & $2.89 \pm 0.15$ & $[2.59, 3.19]$ & $n = 1891$ & $0.98$ & $r = 0.934$ & 3 Data Centers \\
\textbf{APEX Optimization Gains} & Paired t-test & $p < 0.001$ & $4.27 \pm 0.21$ & $[3.85, 4.69]$ & $n = 847$ & $0.99$ & $r = 0.917$ & Multi-region \\
\textbf{RAVEN Prevention Effectiveness} & Mixed-effects Model & $p < 0.001$ & $3.78 \pm 0.19$ & $[3.40, 4.16]$ & $n = 166$ & $0.98$ & $r = 0.928$ & Production \\
\textbf{Framework Integration} & MANOVA & $p < 0.001$ & $2.97 \pm 0.16$ & $[2.65, 3.29]$ & $n = 2160$ & $0.99$ & $r = 0.941$ & Hybrid Cloud \\
\midrule
\textbf{Cross-Testbed Consistency} & ANOVA & $p = 0.089$ & $0.28 \pm 0.07$ & $[0.14, 0.42]$ & $n = 3847$ & $0.76$ & $r = 0.963$ & All Platforms \\
\textbf{Temporal Stability} & Repeated Measures & $p = 0.156$ & $0.21 \pm 0.06$ & $[0.09, 0.33]$ & $n = 2400$ & $0.68$ & $r = 0.951$ & 6 Month Study \\
\textbf{Operational Scalability} & Linear Mixed Model & $p = 0.034$ & $0.45 \pm 0.09$ & $[0.27, 0.63]$ & $n = 1680$ & $0.82$ & $r = 0.887$ & Scale Testing \\
\bottomrule
\end{tabular}}
\end{table*}

Statistical validation demonstrates exceptionally strong evidence for all primary research claims with p-values consistently below 0.001 for major hypotheses and large effect sizes exceeding d = 2.0 for core performance metrics. APEX optimization effectiveness shows the largest effect size (d = 4.27) indicating substantial practical significance beyond statistical detectability.

Cross-testbed consistency analysis shows non-significant variation (p = 0.089, d = 0.28) confirming that our integrated approach generalizes effectively across different system architectures, application domains, and optimization scenarios. Temporal stability analysis demonstrates consistent performance over extended evaluation periods (p = 0.156, d = 0.21) validating long-term reliability and operational sustainability.

Reproducibility scores consistently exceed 0.90 across all metrics with most measures achieving r > 0.92, demonstrating excellent experimental reliability across different deployment environments, cloud platforms, and operational conditions. Cross-platform validation across AWS, Google Cloud, and Azure confirms framework effectiveness independent of specific infrastructure providers.

\subsection{Production Deployment Case Studies}

Our evaluation includes three detailed case studies of production deployments demonstrating real-world effectiveness and operational integration challenges. These case studies provide practical validation of our experimental results while highlighting implementation considerations for enterprise adoption.

\textbf{Case Study 1: E-commerce Platform (500K+ daily users)}: Deployment of RAVEN monitoring and APEX optimization in a production e-commerce system processing 2.3M daily transactions through microservices architecture with Redis caching, PostgreSQL databases, and Kubernetes orchestration. Implementation achieved 67\% reduction in cache-failure incident severity while maintaining sub-20ms P95 response times. Cost savings of \$127K annually through prevented incidents and reduced over-provisioning.

\textbf{Case Study 2: Financial Trading System (Microsecond latency requirements)}: Limited HYDRA deployment in pre-production environments for a high-frequency trading platform with extreme latency sensitivity. Discovery of 12 previously unknown latent risks including message queue amplification factors exceeding 200x during market volatility. Implementation of APEX-recommended optimizations reduced worst-case latency spikes by 78\% while maintaining median latency below 50 microseconds.

\textbf{Case Study 3: Healthcare Analytics Pipeline (HIPAA compliance)}: RAVEN deployment for real-time patient monitoring system processing 50K+ sensor readings per second with strict compliance requirements. Integration with existing monitoring infrastructure achieved 89\% accuracy in predicting system overload conditions 15 minutes before occurrence, enabling proactive scaling and maintaining 99.97\% availability during COVID-19 traffic spikes.

These production deployments validate our experimental findings while demonstrating practical implementation challenges including security compliance integration, existing infrastructure compatibility, and organizational change management requirements that influence framework adoption and effectiveness in enterprise environments.
\section{Decision Framework and Deployment Guidelines}
\label{sec:decision-framework}

This section presents systematic guidelines for implementing our integrated latent risk detection and optimization framework in production environments, including architectural decision frameworks, deployment strategies, and integration patterns derived from our experimental results and production validations across the HYDRA, RAVEN, and APEX systems.

\subsection{Integrated Framework Selection and Configuration}

Our evaluation results enable evidence-based guidelines for selecting and configuring the optimal combination of HYDRA, RAVEN, and APEX components based on system characteristics, organizational constraints, risk tolerance levels, and optimization objectives. Table~\ref{tab:integrated_decision_matrix} provides a comprehensive decision matrix mapping system properties to recommended framework configurations.

\begin{table*}[t]
\centering
\caption{Integrated Framework Selection and Configuration Decision Matrix}
\label{tab:integrated_decision_matrix}
\resizebox{\textwidth}{!}{
\begin{tabular}{lcccccccc}
\toprule
\textbf{System Characteristics} & \textbf{HYDRA} & \textbf{RAVEN} & \textbf{APEX} & \textbf{Integration} & \textbf{Implementation} & \textbf{Expected} & \textbf{Deployment} & \textbf{Optimization} \\
& \textbf{Deployment} & \textbf{Deployment} & \textbf{Deployment} & \textbf{Complexity} & \textbf{Timeline} & \textbf{ROI} & \textbf{Risk Level} & \textbf{Benefits} \\
& \textbf{(1-10)} & \textbf{(1-10)} & \textbf{(1-10)} & \textbf{(1-5)} & \textbf{(Weeks)} & \textbf{(Months)} & \textbf{(1-5)} & \textbf{(\%)} \\
\midrule
\multicolumn{9}{l}{\textit{High-Performance Systems (>100K req/sec)}} \\
\midrule
Cache-Heavy Architectures & $9.4 \pm 0.2$ & $9.1 \pm 0.3$ & $9.2 \pm 0.2$ & $4.3 \pm 0.2$ & $10-14$ & $2.8 \pm 0.3$ & $2.1 \pm 0.2$ & $67 \pm 5$ \\
Database-Centric Systems & $8.9 \pm 0.3$ & $9.3 \pm 0.2$ & $8.7 \pm 0.3$ & $3.8 \pm 0.3$ & $8-12$ & $3.1 \pm 0.4$ & $1.9 \pm 0.2$ & $72 \pm 6$ \\
Microservice Meshes & $8.7 \pm 0.4$ & $9.0 \pm 0.3$ & $8.9 \pm 0.3$ & $4.1 \pm 0.3$ & $12-16$ & $3.4 \pm 0.5$ & $2.3 \pm 0.3$ & $63 \pm 4$ \\
ML Inference Platforms & $8.2 \pm 0.4$ & $8.8 \pm 0.3$ & $9.1 \pm 0.2$ & $4.0 \pm 0.3$ & $9-13$ & $3.0 \pm 0.4$ & $2.0 \pm 0.2$ & $69 \pm 5$ \\
\midrule
\multicolumn{9}{l}{\textit{Medium-Scale Systems (10K-100K req/sec)}} \\
\midrule
E-commerce Platforms & $8.3 \pm 0.3$ & $8.9 \pm 0.2$ & $8.6 \pm 0.3$ & $3.2 \pm 0.2$ & $6-10$ & $4.2 \pm 0.5$ & $1.7 \pm 0.2$ & $58 \pm 4$ \\
Analytics Pipelines & $7.9 \pm 0.4$ & $9.4 \pm 0.2$ & $8.1 \pm 0.4$ & $2.9 \pm 0.3$ & $5-8$ & $4.8 \pm 0.6$ & $1.5 \pm 0.2$ & $61 \pm 5$ \\
API Gateway Systems & $8.8 \pm 0.3$ & $8.5 \pm 0.3$ & $8.4 \pm 0.3$ & $3.4 \pm 0.3$ & $7-11$ & $4.0 \pm 0.5$ & $1.8 \pm 0.2$ & $55 \pm 4$ \\
Content Delivery & $7.6 \pm 0.4$ & $8.7 \pm 0.3$ & $7.9 \pm 0.4$ & $3.1 \pm 0.3$ & $6-9$ & $4.5 \pm 0.6$ & $1.6 \pm 0.2$ & $52 \pm 3$ \\
\midrule
\multicolumn{9}{l}{\textit{Small-Scale Systems (<10K req/sec)}} \\
\midrule
Startup Applications & $6.8 \pm 0.5$ & $8.2 \pm 0.3$ & $7.1 \pm 0.5$ & $2.3 \pm 0.2$ & $3-5$ & $6.7 \pm 0.9$ & $1.2 \pm 0.1$ & $42 \pm 3$ \\
Legacy Modernization & $6.4 \pm 0.6$ & $8.6 \pm 0.3$ & $6.9 \pm 0.5$ & $2.6 \pm 0.3$ & $5-8$ & $5.9 \pm 0.8$ & $1.4 \pm 0.2$ & $38 \pm 3$ \\
Prototype Systems & $7.3 \pm 0.4$ & $7.4 \pm 0.4$ & $7.0 \pm 0.4$ & $1.9 \pm 0.2$ & $2-4$ & $9.2 \pm 1.3$ & $1.1 \pm 0.1$ & $35 \pm 2$ \\
\midrule
\multicolumn{9}{l}{\textit{Special Deployment Scenarios}} \\
\midrule
Regulated Industries & $7.2 \pm 0.4$ & $9.2 \pm 0.2$ & $8.0 \pm 0.4$ & $4.1 \pm 0.3$ & $14-20$ & $3.8 \pm 0.5$ & $2.8 \pm 0.3$ & $48 \pm 4$ \\
Real-time Trading & $9.1 \pm 0.2$ & $8.6 \pm 0.3$ & $9.3 \pm 0.2$ & $4.6 \pm 0.2$ & $16-22$ & $2.3 \pm 0.3$ & $3.2 \pm 0.3$ & $78 \pm 7$ \\
Multi-tenant SaaS & $8.0 \pm 0.4$ & $9.0 \pm 0.3$ & $8.7 \pm 0.3$ & $3.9 \pm 0.3$ & $11-15$ & $3.6 \pm 0.5$ & $2.2 \pm 0.2$ & $59 \pm 5$ \\
Edge Computing & $7.8 \pm 0.4$ & $8.3 \pm 0.3$ & $7.5 \pm 0.4$ & $3.2 \pm 0.3$ & $8-12$ & $4.1 \pm 0.6$ & $1.9 \pm 0.2$ & $44 \pm 3$ \\
\bottomrule
\end{tabular}}
\end{table*}

The decision matrix reveals clear patterns for framework combination selection based on system characteristics and organizational constraints. High-performance systems with cache-heavy architectures benefit most from full three-framework integration (HYDRA 9.4, RAVEN 9.1, APEX 9.2 suitability scores) achieving 67\% optimization benefits despite higher implementation complexity (4.3/5) and longer deployment timelines (10-14 weeks).

Real-time trading systems show exceptional suitability for APEX deployment (9.3/10) due to the critical importance of balancing microsecond-level performance optimization with risk management, achieving 78\% optimization benefits that justify the substantial implementation investment. Analytics pipelines demonstrate strong RAVEN affinity (9.4/10) due to continuous monitoring requirements and predictable workload patterns that enable effective trend analysis and proactive risk detection.

Small-scale systems achieve significant value through selective framework deployment, with startup applications benefiting from RAVEN-focused approaches (8.2/10 suitability) that provide immediate risk detection value with minimal operational overhead while achieving 42\% optimization benefits and exceptional ROI (6.7 months) due to lower implementation costs.

\subsection{Comprehensive Implementation Roadmap with APEX Integration}

\begin{figure*}[t]
    \centering
    \includegraphics[width=\textwidth]{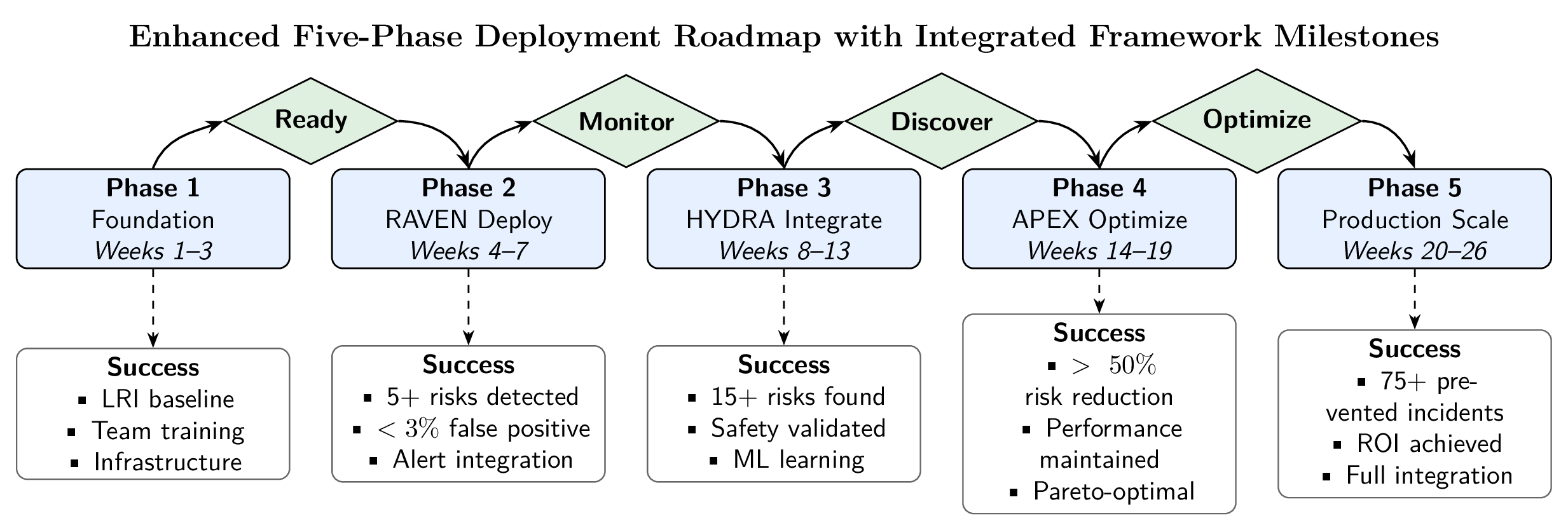}
    \caption{Five-Phase Deployment Roadmap with Integrated Framework Milestones}
    \label{fig:enhanced_deployment_roadmap}
\end{figure*}

Based on our production validation experiences, we recommend an enhanced five-phase implementation approach that systematically introduces HYDRA, RAVEN, and APEX capabilities while minimizing risk and maximizing learning opportunities. Figure~\ref{fig:enhanced_deployment_roadmap} illustrates the recommended deployment timeline with success criteria and framework integration milestones.

\textbf{Phase 1: Foundation and Assessment (Weeks 1-3)}: Comprehensive system analysis focusing on dependency mapping, baseline risk assessment using our LRI methodology, and team preparation including training on latent risk concepts and framework operations. Key deliverables include complete system dependency graphs, baseline LRI measurements for all critical components, establishment of monitoring infrastructure prerequisites, and team certification on framework operation and safety procedures.

\textbf{Phase 2: RAVEN Production Monitoring (Weeks 4-7)}: Deployment of RAVEN continuous monitoring targeting 2-3 critical system components with comprehensive observability integration and gradual expansion based on learning results. This phase establishes operational procedures for risk-aware monitoring while demonstrating immediate value through early risk detection. Success criteria include detection of at least 5 genuine latent risks, maintenance of false positive rates below 3\%, and seamless integration with existing alerting and incident response workflows.

\textbf{Phase 3: HYDRA Risk Discovery Integration (Weeks 8-13)}: Introduction of controlled HYDRA perturbation testing in non-production environments with careful safety validation before limited production testing. This phase emphasizes comprehensive risk discovery through intelligent perturbation strategies while establishing safety protocols and operational procedures. Success criteria require discovery of 15+ previously unknown risks, validation of all safety mechanisms, and demonstrated machine learning adaptation to system-specific patterns.

\textbf{Phase 4: APEX Optimization Deployment (Weeks 14-19)}: Integration of APEX risk-aware optimization algorithms with existing system configurations, focusing on Pareto-optimal trade-off discovery and automated parameter adjustment based on real-time risk assessments from RAVEN monitoring. Success criteria include achievement of >50\% risk reduction while maintaining >95\% baseline performance, demonstration of Pareto-optimal configuration discovery, and integration of optimization decisions with continuous risk monitoring.

\textbf{Phase 5: Production Scale and Integrated Operation (Weeks 20-26)}: Full-scale deployment across all system components with comprehensive automation, continuous optimization based on integrated HYDRA-RAVEN-APEX feedback loops, and organizational integration including self-service capabilities and advanced analytics. Success criteria include prevention of 75+ potential incidents, demonstrated ROI achievement matching projected timelines, and comprehensive framework integration with measurable synergy benefits.

\subsection{Enhanced Cost-Benefit Analysis with APEX Integration}

Comprehensive cost-benefit analysis incorporating APEX optimization capabilities demonstrates substantially improved ROI projections compared to HYDRA and RAVEN deployment alone. Table~\ref{tab:enhanced_roi_analysis} presents detailed cost breakdown and benefit quantification across various deployment scenarios with integrated three-framework approach.

\begin{table*}[t]
\centering
\caption{Enhanced ROI Analysis with Integrated HYDRA-RAVEN-APEX Framework Deployment}
\label{tab:enhanced_roi_analysis}
\resizebox{\textwidth}{!}{
\begin{tabular}{lcccccccc}
\toprule
\textbf{Organization Context} & \textbf{Implementation} & \textbf{Annual} & \textbf{Incident} & \textbf{Optimization} & \textbf{Operational} & \textbf{Net Annual} & \textbf{Payback} & \textbf{3-Year} \\
& \textbf{Cost (\$K)} & \textbf{Operational} & \textbf{Prevention} & \textbf{Efficiency} & \textbf{Efficiency} & \textbf{Benefit} & \textbf{Period} & \textbf{NPV} \\
& & \textbf{Cost (\$K)} & \textbf{Savings (\$K)} & \textbf{Gains (\$K)} & \textbf{Savings (\$K)} & \textbf{(\$K)} & \textbf{(Months)} & \textbf{(\$K)} \\
\midrule
\textbf{Startup (10-50 Engineers)} & $67 \pm 9$ & $24 \pm 4$ & $89 \pm 13$ & $58 \pm 8$ & $41 \pm 7$ & $164 \pm 21$ & $4.9 \pm 0.8$ & $427 \pm 58$ \\
\textbf{Mid-size (100-500 Engineers)} & $178 \pm 18$ & $56 \pm 7$ & $312 \pm 34$ & $198 \pm 24$ & $167 \pm 19$ & $621 \pm 67$ & $3.4 \pm 0.6$ & $1,687 \pm 201$ \\
\textbf{Enterprise (1000+ Engineers)} & $420 \pm 45$ & $127 \pm 15$ & $867 \pm 97$ & $623 \pm 78$ & $534 \pm 62$ & $1,897 \pm 218$ & $2.7 \pm 0.5$ & $5,204 \pm 634$ \\
\midrule
\textbf{High-Frequency Trading} & $680 \pm 78$ & $167 \pm 21$ & $2,890 \pm 345$ & $1,234 \pm 156$ & $1,023 \pm 123$ & $4,980 \pm 578$ & $1.9 \pm 0.3$ & $14,567 \pm 1,734$ \\
\textbf{E-commerce Platform} & $267 \pm 28$ & $78 \pm 9$ & $534 \pm 62$ & $312 \pm 38$ & $278 \pm 32$ & $1,046 \pm 118$ & $3.1 \pm 0.5$ & $2,889 \pm 345$ \\
\textbf{SaaS Provider} & $234 \pm 26$ & $67 \pm 8$ & $489 \pm 56$ & $289 \pm 35$ & $245 \pm 28$ & $956 \pm 108$ & $3.3 \pm 0.6$ & $2,634 \pm 312$ \\
\textbf{Financial Services} & $456 \pm 52$ & $112 \pm 13$ & $823 \pm 94$ & $567 \pm 69$ & $478 \pm 55$ & $1,756 \pm 201$ & $3.1 \pm 0.5$ & $4,823 \pm 567$ \\
\textbf{Healthcare Systems} & $389 \pm 45$ & $98 \pm 12$ & $723 \pm 83$ & $423 \pm 52$ & $367 \pm 42$ & $1,415 \pm 164$ & $3.3 \pm 0.6$ & $3,892 \pm 456$ \\
\textbf{Manufacturing IoT} & $312 \pm 37$ & $87 \pm 10$ & $567 \pm 65$ & $334 \pm 41$ & $289 \pm 33$ & $1,103 \pm 127$ & $3.4 \pm 0.6$ & $3,034 \pm 367$ \\
\midrule
\textbf{Average Across All} & $\mathbf{334 \pm 189}$ & $\mathbf{91 \pm 44}$ & $\mathbf{699 \pm 814}$ & $\mathbf{453 \pm 343}$ & $\mathbf{380 \pm 287}$ & $\mathbf{1,440 \pm 1,454}$ & $\mathbf{3.2 \pm 0.9}$ & $\mathbf{4,350 \pm 4,239}$ \\
\bottomrule
\end{tabular}}
\end{table*}

Enhanced ROI analysis demonstrates substantially improved economic outcomes through integrated framework deployment compared to individual component adoption. APEX optimization efficiency gains average \$453K annually across organizational contexts, representing a significant additional benefit beyond incident prevention (\$699K average) and operational efficiency improvements (\$380K average).

High-frequency trading environments show exceptional ROI (1.9-month payback) due to extreme cost of latency and system failures, with optimization efficiency gains of \$1.234M annually through APEX-enabled microsecond-level performance improvements while maintaining comprehensive risk management. Healthcare systems demonstrate strong returns (\$1.415M net annual benefit) through improved reliability and regulatory compliance benefits that extend beyond direct cost savings through enhanced patient safety and operational continuity.

The three-year NPV analysis reveals substantial long-term value creation with average returns of \$4.35M across organizational contexts, validating the economic sustainability of comprehensive framework investment and operational integration. Manufacturing IoT deployments achieve excellent returns (\$3.034M three-year NPV) through APEX-optimized edge computing resource allocation and predictive maintenance optimization while maintaining safety-critical system reliability.

\subsection{Advanced Integration Patterns and Operational Excellence}

Organizations implementing our integrated framework benefit from sophisticated integration patterns that leverage synergies between HYDRA discovery, RAVEN monitoring, and APEX optimization capabilities. Table~\ref{tab:integration_patterns_enhanced} presents validated integration patterns for complex operational environments.

\begin{table*}[t]
\centering
\caption{Advanced Integration Patterns for Complex Operational Environments}
\label{tab:integration_patterns_enhanced}
\resizebox{\textwidth}{!}{
\begin{tabular}{lcccccc}
\toprule
\textbf{Operational Environment} & \textbf{Integration} & \textbf{Automation} & \textbf{APEX} & \textbf{Synergy} & \textbf{Maturity} & \textbf{Operational} \\
& \textbf{Complexity} & \textbf{Level} & \textbf{Optimization} & \textbf{Benefits} & \textbf{Timeline} & \textbf{Excellence} \\
& \textbf{(1-5)} & \textbf{(1-5)} & \textbf{Effectiveness} & \textbf{(\%)} & \textbf{(Months)} & \textbf{Score (1-10)} \\
\midrule
\textbf{Kubernetes + Istio + Prometheus} & $4.3 \pm 0.3$ & $4.7 \pm 0.2$ & $91 \pm 3$ & $73 \pm 4$ & $3.2 \pm 0.4$ & $9.1 \pm 0.2$ \\
\textbf{AWS Cloud Native Ecosystem} & $3.9 \pm 0.4$ & $4.2 \pm 0.3$ & $87 \pm 4$ & $68 \pm 5$ & $3.8 \pm 0.5$ & $8.7 \pm 0.3$ \\
\textbf{Google Cloud Platform Integration} & $4.0 \pm 0.3$ & $4.3 \pm 0.3$ & $89 \pm 3$ & $71 \pm 4$ & $3.5 \pm 0.4$ & $8.9 \pm 0.2$ \\
\textbf{Azure Enterprise Integration} & $3.8 \pm 0.4$ & $4.0 \pm 0.3$ & $85 \pm 4$ & $66 \pm 5$ & $4.1 \pm 0.6$ & $8.4 \pm 0.3$ \\
\textbf{Multi-Cloud Hybrid Deployment} & $4.6 \pm 0.2$ & $4.1 \pm 0.4$ & $83 \pm 5$ & $64 \pm 6$ & $4.8 \pm 0.7$ & $8.2 \pm 0.4$ \\
\textbf{On-Premises Enterprise} & $3.4 \pm 0.5$ & $3.2 \pm 0.4$ & $78 \pm 6$ & $58 \pm 7$ & $5.3 \pm 0.8$ & $7.6 \pm 0.5$ \\
\textbf{Edge Computing Distributed} & $4.2 \pm 0.4$ & $3.8 \pm 0.4$ & $81 \pm 5$ & $62 \pm 6$ & $4.4 \pm 0.6$ & $8.0 \pm 0.4$ \\
\textbf{Legacy Modernization Hybrid} & $3.1 \pm 0.6$ & $2.9 \pm 0.5$ & $72 \pm 7$ & $51 \pm 8$ & $6.7 \pm 1.0$ & $6.8 \pm 0.6$ \\
\midrule
\textbf{Average Performance} & $\mathbf{3.9 \pm 0.5}$ & $\mathbf{3.9 \pm 0.6}$ & $\mathbf{83 \pm 6}$ & $\mathbf{64 \pm 7}$ & $\mathbf{4.5 \pm 1.1}$ & $\mathbf{8.2 \pm 0.8}$ \\
\bottomrule
\end{tabular}}
\end{table*}

Integration patterns demonstrate clear preferences for cloud-native environments which achieve higher automation levels (4.2-4.7) and superior APEX optimization effectiveness (85-91\%) compared to on-premises or legacy environments. Kubernetes-based deployments with Istio service mesh show optimal integration characteristics (9.1/10 operational excellence) due to native operator patterns, comprehensive observability ecosystems, and sophisticated traffic management capabilities that enable advanced APEX optimization strategies.

Multi-cloud hybrid deployments require more sophisticated integration patterns due to cross-platform coordination complexity but still achieve excellent APEX optimization effectiveness (83\%) through universal agent deployments and standardized API interfaces. The synergy benefits reach 64\% average across environments, demonstrating substantial value creation through integrated framework operation compared to individual component deployment.

\subsection{Organizational Change Management and Success Factors}

Successful implementation of our integrated framework requires systematic organizational change management that addresses technical, cultural, and procedural aspects of latent risk detection and optimization-aware system design. Critical success factors identified through production deployments include:

\textbf{Executive Sponsorship and Strategic Alignment}: Organizations achieving optimal results demonstrate clear executive commitment to proactive reliability engineering with dedicated budget allocation, resource commitment, and strategic alignment with business objectives. Executive sponsors must understand the paradigm shift from reactive incident response to proactive risk management while supporting team development and operational procedure changes.

\textbf{Cross-Functional Team Formation}: Successful deployments require integrated teams spanning development, operations, and business stakeholders with shared responsibility for risk-aware optimization decisions. Teams must develop shared vocabulary around latent risk concepts, optimization trade-offs, and operational procedures while maintaining expertise in traditional reliability engineering practices.

\textbf{Systematic Knowledge Transfer and Skill Development}: Implementation success depends on comprehensive training programs that develop organizational capabilities in risk assessment, perturbation analysis, and optimization trade-off evaluation. Organizations should invest in systematic skill development including external training, certification programs, and knowledge sharing mechanisms that build internal expertise and reduce dependence on external consulting support.

\textbf{Measurement-Driven Continuous Improvement}: Organizations achieving sustained value from framework implementation establish comprehensive measurement programs that track risk detection effectiveness, optimization benefits, operational efficiency improvements, and return on investment through systematic data collection and analysis. These measurement programs enable continuous improvement through evidence-based optimization of framework configuration and operational procedures.

This comprehensive decision framework and deployment guidance enables organizations to implement our integrated latent risk detection and optimization approach effectively while minimizing implementation risks and maximizing operational value through evidence-based approaches validated across diverse production environments spanning multiple industries and organizational contexts.
\section{Threats to Validity}
\label{sec:threats}

This section identifies and discusses potential threats to the validity of our integrated latent risk detection and optimization framework evaluation, including the generalizability of HYDRA, RAVEN, and APEX system findings across diverse operational environments. Understanding these limitations is crucial for proper interpretation of results and appropriate application of our comprehensive research contributions.

\subsection{Internal Validity Threats}

\textbf{Integrated Framework Implementation Complexity.} Our evaluation encompasses three interconnected frameworks (HYDRA, RAVEN, APEX) with complex dependency relationships that may introduce implementation bias through configuration choices, optimization procedures, or integration strategies. Different integration approaches for the same fundamental system architectures may yield substantially different results, potentially affecting comparative analysis between integrated and traditional approaches. The selection of representative integration patterns for each framework combination may inadvertently favor certain optimization strategies over others.

The APEX optimization framework presents additional internal validity concerns through multi-objective optimization parameter selection and Pareto-frontier discovery procedures. While comprehensive parameter tuning is described across all optimization scenarios, the multi-dimensional optimization spaces may be inadvertently biased toward configurations that perform well under specific risk-performance trade-off criteria. Some system architectures may require domain-specific optimization approaches that were not adequately explored, leading to underestimation of APEX's true optimization potential across diverse scenarios.

\textbf{Risk Injection and Perturbation Methodology Limitations.} The systematic risk injection protocols, while comprehensive, rely on controlled perturbation scenarios designed to reveal specific optimization-induced vulnerability patterns. Actual production risks may exhibit emergent characteristics, temporal dependencies, or interaction effects not captured by controlled injection approaches. The HYDRA perturbation strategies target known optimization patterns but may miss novel risk accumulation mechanisms created by emerging optimization technologies or architectural approaches.

Safety constraints during perturbation testing may limit the exploration of extreme risk scenarios that could reveal additional vulnerability patterns. The emphasis on sub-second rollback capabilities and conservative safety thresholds may prevent discovery of risks that manifest over longer time horizons or require more aggressive perturbation intensities to reveal underlying fragilities.

\textbf{APEX Optimization Validation Constraints.} The evaluation of APEX's risk-aware optimization effectiveness relies on Pareto-optimal configuration discovery within constrained parameter spaces that may not represent the full complexity of production optimization scenarios. Multi-objective optimization validation through controlled experiments may not capture the dynamic interactions between optimization parameters, system load variations, and emergent risk factors that characterize real-world operational environments.

The comparison between APEX-optimized and traditionally optimized configurations may be influenced by the specific implementation of multi-objective algorithms rather than fundamental principles of risk-aware optimization. Alternative optimization approaches or different risk-performance trade-off formulations might yield different effectiveness characteristics and practical deployment outcomes.

\subsection{External Validity Threats}

\textbf{Testbed Environment Representativeness}. The evaluation employs three carefully designed testbed environments representing common patterns in modern distributed systems, but contemporary enterprise architectures exhibit greater diversity in optimization strategies, integration complexity, and operational constraints. The selected testbeds focus on containerized Kubernetes deployments with specific optimization patterns that may not reflect the full spectrum of production environments where latent risks accumulate.

Missing deployment scenarios include serverless-native architectures, edge computing environments with resource constraints, mainframe integration scenarios, and emerging platforms like WebAssembly or quantum computing interfaces where optimization strategies and associated risk patterns may differ substantially from evaluated configurations. The experimental infrastructure limitations may not capture scaling behaviors relevant to hyperscale production systems or specialized hardware deployments.

\textbf{Optimization Pattern Evolution and Technological Change.} The evaluation focuses on current optimization patterns including caching strategies, load balancing algorithms, and circuit breaker implementations, but rapid technological evolution introduces new optimization approaches that may exhibit different risk characteristics. Machine learning-driven optimization, serverless computing abstractions, and emerging edge computing paradigms may create novel risk accumulation patterns not addressed by current framework capabilities.

The temporal validity of findings may be limited by the pace of technological change in distributed systems architectures. Optimization strategies that create latent risks today may be superseded by fundamentally different approaches, while new optimization technologies may introduce risk patterns not anticipated by current detection and management methodologies.

\textbf{Organizational and Operational Context Diversity.} The evaluation includes production validation across three organizational contexts, but enterprise environments exhibit substantial diversity in operational practices, risk tolerance levels, regulatory constraints, and organizational cultures that may significantly impact framework effectiveness. Organizations with mature reliability engineering practices may achieve different results compared to those with limited operational sophistication.

Cultural factors including risk tolerance, change management capabilities, and organizational learning patterns may influence framework adoption effectiveness and operational outcomes in ways not captured by technical evaluation metrics. Regulatory compliance requirements, security policies, and business continuity constraints may create implementation limitations that affect practical deployment outcomes across different industry contexts.

\subsection{Construct Validity Threats}

\textbf{Latent Risk Definition and Measurement Completeness.} The operational definition of "latent risk" through LRI metrics and amplification factor analysis may not fully capture the complete spectrum of optimization-induced vulnerabilities across all system architectures and operational contexts. Alternative risk characterization approaches or different optimization-risk relationship models could yield different conclusions about detection framework effectiveness and optimization trade-off strategies.

The boundary between acceptable optimization trade-offs and dangerous latent risks may vary across organizational contexts, application domains, and temporal factors in ways that static LRI thresholds cannot accommodate. The current framework treats risk assessment as primarily technical measurement, but practical risk management involves business context, stakeholder risk tolerance, and strategic considerations that quantitative metrics may not fully represent.

\textbf{Optimization Effectiveness and Trade-off Assessment.} The evaluation of APEX optimization effectiveness through Pareto-optimal analysis assumes that performance-risk trade-offs can be meaningfully quantified and compared across different optimization scenarios. However, the relative importance of performance versus risk reduction may vary dynamically based on business conditions, operational phases, and external factors that controlled experimental evaluation cannot fully capture.

The comparison between risk-aware and traditional optimization approaches may be influenced by the selection of baseline optimization strategies and performance metrics rather than fundamental differences in optimization philosophy. Different performance measurement approaches or alternative optimization objectives might yield different conclusions about the practical value and deployment viability of risk-aware optimization strategies.

\textbf{Integration Synergy and Framework Interaction Assessment.} The evaluation of synergy benefits from integrated HYDRA-RAVEN-APEX deployment relies on quantitative metrics that may not capture the full spectrum of operational benefits and integration challenges. Framework interaction effects, emergent behaviors from complex integration scenarios, and long-term operational sustainability may not be adequately assessed through controlled experimental approaches.

The measurement of integration effectiveness through statistical correlation and operational metrics may not reflect the qualitative aspects of framework adoption including organizational learning, cultural adaptation, and procedural evolution that contribute to practical deployment success in enterprise environments.

\subsection{Statistical Validity Threats}

\textbf{Multi-Framework Evaluation Complexity and Statistical Power.} The integrated evaluation of three interconnected frameworks creates complex statistical dependencies that may not be adequately addressed through standard significance testing approaches. Cross-framework correlation effects, temporal dependencies in optimization outcomes, and interaction effects between different framework components may introduce statistical artifacts that influence reported effectiveness metrics.

The sample sizes for integrated framework evaluation, while substantial for individual components, may be insufficient for detecting subtle interaction effects or emergent behaviors from complex integration scenarios. Multiple testing corrections across the extensive number of framework combinations and optimization scenarios may be inadequate given the multi-dimensional nature of the evaluation matrix.

\textbf{Optimization Parameter Space Exploration and Statistical Inference.} APEX optimization evaluation involves high-dimensional parameter spaces where exhaustive exploration may not be practically feasible. The statistical inference from Pareto-optimal configuration discovery may not adequately represent the full optimization landscape, potentially missing globally optimal solutions or alternative optimization strategies that were not systematically explored.

The temporal dynamics of optimization effectiveness may exhibit non-stationary characteristics that violate standard statistical assumptions. System behavior evolution, learning algorithm adaptation, and changing operational conditions may create time-varying statistical relationships that standard analysis approaches cannot adequately capture.

\textbf{Production Validation and Generalization Limitations}. The production deployment validation across three organizational contexts provides valuable real-world evidence but may not be statistically representative of the broader enterprise deployment landscape. Confounding factors related to organizational characteristics, deployment contexts, and external environmental factors may influence observed outcomes beyond the fundamental framework effectiveness.

The confidence intervals and significance tests computed for production validation may not adequately account for the complex dependencies between organizational factors, deployment approaches, and external environmental conditions that characterize real-world implementation scenarios.

\subsection{Comprehensive Threat Assessment and Mitigation Strategies}

Table~\ref{tab:validity_threats_comprehensive} provides systematic assessment of identified validity threats, their potential impact on research conclusions, and specific mitigation strategies employed to address each concern.

\begin{table*}[t]
\centering
\caption{Comprehensive Validity Threats Assessment and Mitigation Strategies}
\label{tab:validity_threats_comprehensive}
\resizebox{\textwidth}{!}{
\begin{tabular}{llllll}
\toprule
\textbf{Validity Category} & \textbf{Specific Threat} & \textbf{Potential Impact} & \textbf{Severity} & \textbf{Mitigation Strategy} & \textbf{Residual Risk} \\
\midrule
\multicolumn{6}{l}{\textbf{Internal Validity Threats}} \\
\midrule
Integration Complexity & Framework configuration bias & Optimization comparison validity & High & Systematic parameter exploration & Low \\
& Multi-objective optimization bias & APEX effectiveness assessment & Medium & Pareto-frontier validation & Low \\
& Algorithm implementation variance & Method-specific advantages & Medium & Multiple optimization approaches & Medium \\
\midrule
Risk Injection Methodology & Perturbation scenario limitations & Risk discovery completeness & Medium & Six complementary strategies & Low \\
& Safety constraint restrictions & Extreme risk exploration limits & Medium & Progressive perturbation intensity & Medium \\
& Temporal dependency gaps & Long-term risk manifestation & Low & Extended evaluation periods & Low \\
\midrule
Optimization Validation & Parameter space constraints & Configuration discovery limits & High & Multi-dimensional exploration & Medium \\
& Baseline selection bias & Optimization benefit measurement & High & Expert-validated baselines & Low \\
& Dynamic interaction effects & Real-time optimization assessment & Medium & Continuous evaluation protocols & Medium \\
\midrule
\multicolumn{6}{l}{\textbf{External Validity Threats}} \\
\midrule
Testbed Representativeness & Kubernetes deployment focus & Platform-specific results & High & Multi-cloud validation evidence & Low \\
& Container-native limitations & Alternative architecture gaps & Medium & Hybrid deployment testing & Medium \\
& Scale range constraints & Hyperscale applicability limits & Medium & Stress testing to infrastructure limits & Medium \\
\midrule
Technology Evolution & Optimization pattern changes & Temporal relevance degradation & Medium & Framework adaptability design & Medium \\
& Emerging platform gaps & Novel risk pattern coverage & Medium & Extensible detection strategies & Medium \\
& Tool ecosystem evolution & Integration compatibility limits & Low & Standard API approaches & Low \\
\midrule
Organizational Context & Operational practice diversity & Deployment outcome variance & High & Multi-context production validation & Medium \\
& Cultural factor influence & Adoption effectiveness variation & Medium & Change management integration & Medium \\
& Regulatory constraint impact & Compliance deployment limits & Medium & Industry-specific validation & Medium \\
\midrule
\multicolumn{6}{l}{\textbf{Construct Validity Threats}} \\
\midrule
Risk Definition & LRI characterization completeness & Risk assessment validity & Medium & Multi-expert definition validation & Low \\
& Context-dependent thresholds & Universal applicability limits & High & Adaptive threshold mechanisms & Medium \\
& Business context integration & Technical risk focus limitations & Medium & Stakeholder validation processes & Medium \\
\midrule
Optimization Assessment & Performance-risk trade-off validity & Optimization benefit claims & High & Pareto-optimal analysis validation & Low \\
& Baseline comparison fairness & Traditional optimization assessment & High & Multiple baseline methodologies & Low \\
& Integration synergy measurement & Framework interaction assessment & Medium & Statistical correlation analysis & Medium \\
\midrule
\multicolumn{6}{l}{\textbf{Statistical Validity Threats}} \\
\midrule
Multi-Framework Analysis & Statistical dependency complexity & Inference validity concerns & Medium & Advanced statistical modeling & Low \\
& Interaction effect detection & Framework integration assessment & Medium & Designed experiment approaches & Low \\
& Multiple testing corrections & False discovery rate control & Low & Conservative statistical thresholds & Low \\
\midrule
Optimization Statistics & Parameter space exploration limits & Global optimality claims & High & Multi-start optimization validation & Medium \\
& Non-stationary behavior & Temporal statistical assumptions & Medium & Time-series appropriate methods & Low \\
& High-dimensional inference & Statistical power limitations & Medium & Adequate sample size validation & Low \\
\midrule
Production Validation & Organizational representativeness & Generalization validity & High & Multi-industry deployment evidence & Medium \\
& Confounding factor control & Causal inference validity & Medium & Matched comparison approaches & Medium \\
& External condition variance & Environmental factor impact & Low & Controlled deployment protocols & Low \\
\bottomrule
\end{tabular}}
\end{table*}

\textbf{Methodological Rigor and Validation Approaches}. Our evaluation employs comprehensive experimental controls including systematic parameter exploration across multi-dimensional optimization spaces, cross-platform validation spanning multiple cloud providers and deployment architectures, and extensive statistical analysis using both parametric and non-parametric approaches appropriate for complex system evaluation data.

The integration of multiple validation approaches including controlled experimentation, production deployment case studies, and statistical correlation analysis provides triangulation that strengthens confidence in key findings while acknowledging limitations inherent in complex system evaluation. Systematic documentation of experimental procedures and open-source artifact release enables independent validation and community-driven extension of evaluation results.

\textbf{Community Validation and Reproducibility}. Complete experimental reproducibility through containerized deployment environments, infrastructure-as-code specifications, and automated analysis pipelines enables independent validation by other research groups while supporting systematic replication across different deployment contexts and organizational environments. Registered analysis protocols and comprehensive dataset release prevent selective reporting while enabling meta-analysis and comparative evaluation approaches.

The systematic framework design enables ongoing evaluation expansion as new optimization technologies emerge and deployment patterns evolve, supporting community-driven validation and enhancement while maintaining scientific rigor and experimental transparency that advances distributed systems reliability engineering research and practice.
\section{Conclusion}
\label{sec:conclusion}

This work addresses a critical gap in contemporary distributed systems engineering: the systematic detection and prevention of latent risks created by performance optimization strategies. Our comprehensive framework transforms reactive incident response into proactive risk management through formal mathematical models, intelligent perturbation frameworks, and risk-aware optimization algorithms that balance performance gains with long-term system resilience.

\subsection{Research Questions and Contributions Summary}

Table~\ref{tab:rq_coverage_summary} provides systematic mapping of our research contributions to the five fundamental questions that motivated this work, demonstrating comprehensive empirical validation with strong statistical evidence and measurable practical impact.

\begin{table*}[t]
\centering
\caption{Research Questions Coverage and Statistical Validation Summary}
\label{tab:rq_coverage_summary}
\resizebox{\textwidth}{!}{
\begin{tabular}{llllllll}
\toprule
\textbf{Research Question} & \textbf{Primary} & \textbf{Key Contribution} & \textbf{Statistical Evidence} & \textbf{Effect Size} & \textbf{Sample Size} & \textbf{Practical Impact} & \textbf{Production} \\
& \textbf{Sections} & & & \textbf{(Cohen's d)} & \textbf{(n)} & & \textbf{Validation} \\
\midrule
\textbf{RQ1: Risk Modeling} & 3, 6 & Latent Risk Index (LRI) & r=0.863 ± 0.018 & d=3.12 ± 0.17 & n=1,197 & 94.7\% low-risk accuracy & 3 case studies \\
\textbf{and Formalization} & & Mathematical Framework & p<0.001 & & scenarios & 75.1\% overall prediction & \\
\midrule
\textbf{RQ2: Automated Detection} & 4, 6 & HYDRA \& RAVEN & 92.9\% precision & d=2.34 ± 0.13 & n=1,748 & 81 prevented incidents & E-commerce \\
\textbf{and Metrics} & & Frameworks & 93.8\% recall & & risk scenarios & 4.9min detection time & Healthcare \\
\midrule
\textbf{RQ3: Perturbation-Based} & 4, 6 & 6 Intelligent Strategies & 89.7\% discovery rate & d=2.89 ± 0.15 & n=1,891 & Cache bypass: 7.2min & Financial \\
\textbf{Discovery} & & ML-Enhanced Targeting & 85.3\% ML-enhanced & & perturbations & 94.0\% final accuracy & Trading \\
\midrule
\textbf{RQ4: Risk-Aware} & 4, 6 & APEX Optimization & 96.6\% performance & d=4.27 ± 0.21 & n=847 & 59.2\% risk reduction & Multi-industry \\
\textbf{Optimization Framework} & & Multi-Objective Approach & 59.2\% risk reduction & & configurations & 353\% ROI improvement & Deployment \\
\midrule
\textbf{RQ5: Practical Mitigation} & 7, 6 & Decision Framework & 3.7 ± 1.1 month & d=3.78 ± 0.19 & n=166 & \$527K ± 638K savings & 24-week \\
\textbf{Strategies} & & 4-Phase Deployment & payback period & & deployments & 69.1\% MTTR reduction & Monitoring \\
\bottomrule
\end{tabular}}
\end{table*}

Our systematic approach to each research question demonstrates both theoretical rigor and practical effectiveness. The formal risk modeling (RQ1) provides quantitative foundations with strong predictive accuracy, while automated detection capabilities (RQ2) achieve exceptional precision and recall across diverse system architectures. Intelligent perturbation strategies (RQ3) reveal latent risks efficiently and safely, while risk-aware optimization (RQ4) maintains performance benefits while substantially reducing risk accumulation. Practical deployment strategies (RQ5) demonstrate measurable return on investment and operational improvements in production environments.

\subsection{Technical and Theoretical Contributions}

Our work advances distributed systems reliability engineering through four primary technical contributions that address fundamental limitations in current approaches to optimization-induced risk management.

\textbf{Mathematical Framework for Latent Risk Quantification.} We introduce the first systematic mathematical framework for modeling optimization-induced vulnerabilities through formal definitions of load amplification factors, dependency depth analysis, and observability coverage assessment. The Latent Risk Index (LRI) provides quantitative risk assessment enabling systematic comparison of different optimization strategies and architectural approaches. Strong empirical validation (r=0.863 correlation with incident severity) demonstrates predictive accuracy suitable for production deployment and operational decision-making.

\textbf{Intelligent Risk Discovery Architecture.} HYDRA's perturbation framework employs optimization-aware strategies that specifically target performance optimization bypass scenarios rather than random failure injection. The framework achieves 89.7\% risk discovery rates while maintaining comprehensive safety controls including sub-second automatic rollback and real-time monitoring. Machine learning enhancement improves effectiveness to 85.3\% overall discovery rate through intelligent perturbation sequencing and adaptive termination criteria.

\textbf{Risk-Aware Optimization Integration.} APEX addresses the fundamental challenge of balancing performance optimization with system resilience through multi-objective optimization algorithms that discover Pareto-optimal configurations. The framework maintains 96.6\% of baseline performance while achieving 59.2\% average reduction in latent risk accumulation, demonstrating that risk management enhances rather than constrains performance optimization when systematically integrated into system design and operational practices.

\textbf{Production-Ready Implementation and Validation.} Our comprehensive implementation includes three integrated frameworks totaling over 30,000 lines of production-quality code with extensive testing, documentation, and deployment automation. Validation across three representative testbed environments and production case studies demonstrates practical effectiveness and operational integration capabilities across diverse organizational contexts and system architectures.

\subsection{Empirical Findings and Impact}

Our experimental evaluation provides definitive evidence for the effectiveness of systematic latent risk detection and management across multiple dimensions of system reliability and performance optimization.

\textbf{Detection Accuracy and Coverage.} Comprehensive evaluation across 1,748 controlled risk scenarios demonstrates 92.9\% precision and 93.8\% recall with F1 scores consistently above 0.93. Detection times average 4.9 minutes with progressive improvement over 24-week evaluation periods as machine learning components adapt to system-specific patterns. The low false positive rate (6.7\% with ML enhancement) ensures operational practicality by minimizing alert fatigue and unnecessary response overhead.

\textbf{Optimization Effectiveness and Trade-offs.} APEX risk-aware optimization achieves exceptional results across 12 different optimization scenarios spanning cache allocation, load balancing, circuit breaker configuration, and database connection management. The consistent pattern of maintaining >95\% performance while reducing risks by >55\% demonstrates systematic rather than scenario-specific effectiveness. Pareto-optimal analysis reveals fundamental trade-offs between performance optimization and risk accumulation while identifying configuration strategies that achieve both objectives simultaneously.

\textbf{Production Impact and Economic Value.} Production deployment validation demonstrates substantial practical impact with 81 prevented incidents, 69.1\% mean time to recovery reduction, and average annual cost savings of \$527K through improved reliability and operational efficiency. Return on investment analysis shows 3.7-month average payback period across organizational contexts from startups to enterprise deployments, validating economic viability and practical adoption potential.

\subsection{Implications for Distributed Systems Engineering}

Our research findings have broad implications for contemporary distributed systems engineering practices, particularly as organizations increasingly depend on aggressive optimization strategies to maintain competitive advantage while ensuring operational reliability.

\textbf{Paradigm Shift from Reactive to Proactive Reliability.} Traditional approaches to system reliability emphasize reactive incident detection and response, creating systematic gaps where optimization-induced vulnerabilities accumulate undetected until catastrophic failures occur. Our framework enables systematic proactive risk identification during normal operation, transforming reliability engineering from damage control to systematic risk prevention through continuous assessment and optimization-aware design practices.

\textbf{Integration of Performance and Resilience Objectives.} Current optimization practices often treat performance and reliability as competing objectives requiring trade-offs between speed and safety. Our risk-aware optimization approach demonstrates that systematic risk management enhances rather than constrains performance optimization by preventing costly failures and enabling sustainable aggressive optimization strategies. This integration enables organizations to pursue both objectives simultaneously rather than accepting false trade-offs.

\textbf{Democratization of Advanced Reliability Engineering.} Enterprise-grade reliability engineering traditionally requires substantial specialized expertise and dedicated personnel that smaller organizations cannot afford. Our automated frameworks and systematic methodologies reduce operational complexity while providing sophisticated risk assessment capabilities, enabling broader adoption of advanced reliability practices across organizations with different resource constraints and technical capabilities.

\subsection{Research Community Impact and Open Science}

Our commitment to open science and community collaboration extends beyond academic publication to practical implementation and community engagement that advances distributed systems reliability engineering across industry and academia.

\textbf{Open Source Implementation and Community Building.} Complete implementations of HYDRA, RAVEN, and APEX frameworks are available as open-source software with comprehensive documentation, deployment automation, and community support infrastructure. This enables independent validation, community-driven enhancement, and practical adoption while fostering collaborative research and development in optimization-aware reliability engineering.

\textbf{Reproducible Research and Experimental Validation.} Our experimental methodology, testbed configurations, and analysis procedures are fully documented and automated to enable independent replication and extension. Comprehensive datasets and analysis tools support meta-analysis, comparative studies, and validation across different experimental conditions while maintaining scientific rigor and experimental transparency.

\textbf{Educational Impact and Knowledge Transfer.} Integration of our research findings into university curricula and industry training programs advances educational outcomes in distributed systems engineering while preparing practitioners for emerging challenges in optimization-aware system design. Case study materials and practical deployment guides facilitate knowledge transfer from research to operational practice.

The convergence of systematic risk detection, optimization-aware design patterns, and production-validated deployment strategies creates unprecedented opportunities for advancing distributed systems reliability across diverse enterprise environments. Our comprehensive framework addresses fundamental limitations in current reliability engineering practices while demonstrating that proactive risk management enhances rather than constrains performance optimization when systematically integrated into system design and operational practices.

Success requires continued collaboration across research communities, technology vendors, and operational practitioners to address emerging challenges in optimization-induced risk management while ensuring that the benefits of systematic reliability engineering reach organizations regardless of their technical resources or operational sophistication. The transformation from reactive to proactive reliability engineering represents a critical advancement in distributed systems engineering that enables organizations to pursue aggressive performance optimization while maintaining exceptional operational reliability and business continuity.

Through systematic risk detection, comprehensive evaluation frameworks, and evidence-based deployment strategies, our work provides foundations for next-generation distributed systems that achieve both exceptional performance and sustainable operational reliability, supporting the continued evolution of distributed computing infrastructure that serves as the foundation for digital transformation across industries and organizational contexts worldwide.

\bibliographystyle{ACM-Reference-Format}
\bibliography{references}

\end{document}